\documentclass[aps,prb,twocolumn,superscriptaddress,floatfix]{revtex4-2}

\usepackage{dcolumn}
\usepackage{graphicx}
\usepackage{amsfonts}
\usepackage{amsmath,bm,amssymb}
\usepackage{mathtools}
\usepackage{comment}
\usepackage[usenames]{color}
\usepackage{array}
\usepackage[commandnameprefix=always]{changes}
\usepackage[section]{placeins}
\usepackage{hyperref}
\hypersetup{colorlinks=true, linkcolor=blue, urlcolor=blue, citecolor=blue}

\DeclarePairedDelimiterX\braket[2]{\langle}{\rangle}{#1 \delimsize\vert #2}

\DeclareMathOperator{\dd}{\mathrm{d}\!}
\begin{document}

\title{Influence of Magnetic Order on Proximity-Induced Superconductivity in Mn Layers on Nb(110) from First Principles}

\author{Sohair ElMeligy}
\email{sohaire@vt.edu}
\affiliation{Department of Physics, Virginia Tech, Blacksburg, Virginia 24061, USA}
\author{Bal\'azs \'Ujfalussy}
\email{ujfalussy.balazs@wigner.hu}
\affiliation{Wigner Research Centre for Physics, Institute for Solid State Physics and Optics, H-1525 Budapest, Hungary}
\author{Kyungwha Park}
\email{kyungwha@vt.edu}
\affiliation{Department of Physics, Virginia Tech, Blacksburg, Virginia 24061, USA}

\begin{abstract}

 We investigate the influence of magnetic order on the proximity-induced superconducting state in the Mn layers of a Mn-Nb(110) heterostructure by using a first-principles method. For this study, we use the recently developed Bogoliubov-de Gennes (BdG) solver for superconducting heterostructures [Csire et al., Phys. Rev. B 97, 024514 (2018)] within the first-principles calculations based on multiple scattering theory and the screened Korringa-Kohn-Rostoker (SKKR) Green’s function method. In our calculations, we first study the normal-state density of states (DOS) in the single- and double-Mn-layer heterostructures, and calculate the induced magnetic moments in the Nb layers. Next, we compute the momentum-resolved spectral functions in the superconducting state for the heterostructure with a single Mn layer, and find bands crossing the Fermi level within the superconducting (SC) gap. We also study the SC state DOS in the single- and double-Mn-layer heterostructures and compare some of our results with experimental findings, revealing secondary gaps, plateau-like regions, and central V-shaped in-gap states within the bulk SC Nb gap that are magnetic-order-dependent. Finally, we compute the singlet and internally antisymmetric triplet (IAT) order parameters for each layer for both heterostructures, and find an order of magnitude difference in the induced singlet part of the SC order parameter in the Mn layer/s between the FM and AFM cases in favor of the AFM pairing with the maximum still being only 4.44\% of the bulk Nb singlet order parameter value. We also find a negligible induced triplet part, yet comparable to the induced singlet values, indicating some singlet-triplet mixing in the Mn layer/s.
\end{abstract}

\maketitle

\section{Introduction}

Heterostructures made of superconductors and ferromagnetic (FM) or antiferromagnetic (AFM) materials have recently drawn a lot of attention owing to the possibility of realising topological superconductivity and superconducting (SC) spintronics applications \cite{Linder2015, Eschrig_2015, cai2022}. This is allowed by the competitive or cooperative nature of the magnetic materials for realizing spin-singlet and/or spin-triplet Cooper pairs, triplet supercurrents, and increased spin lifetime in the proximity-induced SC state \cite{Bergeret2001, Kraweic2002, Bergeret_2003, Bergeret_2005, Robinson2010, Fritsch_2014, Fritsch_2015, Vezin_2020, Tamura_2023}.

In ferromagnet/superconductor heterostructures, the Cooper pairs in the SC side leak into the FM side, and the exchange field of the FM side generates a spatially varying phase in the SC order parameter. This results in oscillations of the SC critical temperature as a function of the FM film thickness \cite{Demler_1997, Buzdin_2005, Halterman_2001, Kontos_2001, Wu_2012, golubov2025}. Such oscillations were experimentally observed in Gd/Nb bilayer \cite{Jiang1995}, Nb/Gd/Nb trilayers \cite{Khaydukov2018}, Nb/Au/Fe trilayers \cite{Yamazaki2006}, and CoFe/Nb bilayers \cite{Reymond2006}. In antiferromagnet/superconductor heterostructures, the Cooper pairs in the SC side penetrate into the AFM side and the SC critical temperature was observed to decrease as the AFM film thickness increases \cite{Bell2003, Hubener2002, mani_thickness_2015, Chourasia_2023, Fyhn_2023}. On the other hand, a tight-binding model study showed that Neel triplet correlations or the SC critical temperature oscillate as a function of the AFM film thickness \cite{Bobkov2023, Bobkov_2023}, and they were also shown to have a checkerboard or layered Neel triplet type correlations depending on the type of antiferromagnetic ordering \cite{Bobkov_2025, Chourasia_2023}. 

Recently, the proximity-induced SC state has been experimentally studied for Mn atomic layers on a Nb substrate by using scanning tunneling microscopy and spectroscopy. The coexistence of antiferromagnetism and superconductivity was seen in this heterostructure \cite{Lo_Conte_2022} as well as in-gap states within the bulk Nb SC gap. Motivated by this experiment, we apply the recently developed first-principles-based method for SC heterostructures \cite{Csire_2015, Csire2018} to the FM and AFM Mn single and double atomic layers on Nb,  by solving the fully relativistic (Kohn-Sham) Dirac-Bogoliubov-de Gennes (DBdG) equations within the multiple scattering Green's function method (i.e., the screened Korringa-Kohn-Rostoker (SKKR) method). In this method, the band structures of the non-SC layers and SC substrate are taken into account in the SC state. Magnetism, relativity and the proper semi-infinite geometry of the overlayer system are all properly included as well. The Bloch spectral function (BSF), the density of states (DOS), and the charge and anomalous charge densities can be obtained from the single-particle Green's function of the DBdG Hamiltonian, which allows for calculating the singlet and triplet order parameters.

The paper is organized as follows. In Sec. II, we describe the computational steps and parameters used in our calculations, as well as the system of interest (the Mn-Nb(110) heterostructure) in more detail. In Sec. III, we show our results; the normal state DOS, induced magnetic moments in the Nb layers, momentum-resolved Bloch spectral function (BSF) in the SC state, SC state DOS, and the calculated singlet and internally asymmetric triplet (IAT) order parameters for the different layers focusing on the Mn layer/s for the latter. In Sec. IV, we conclude with a summary of our findings.

\section{Computational Details}

\begin{figure}
\centering
\includegraphics[width=\linewidth]{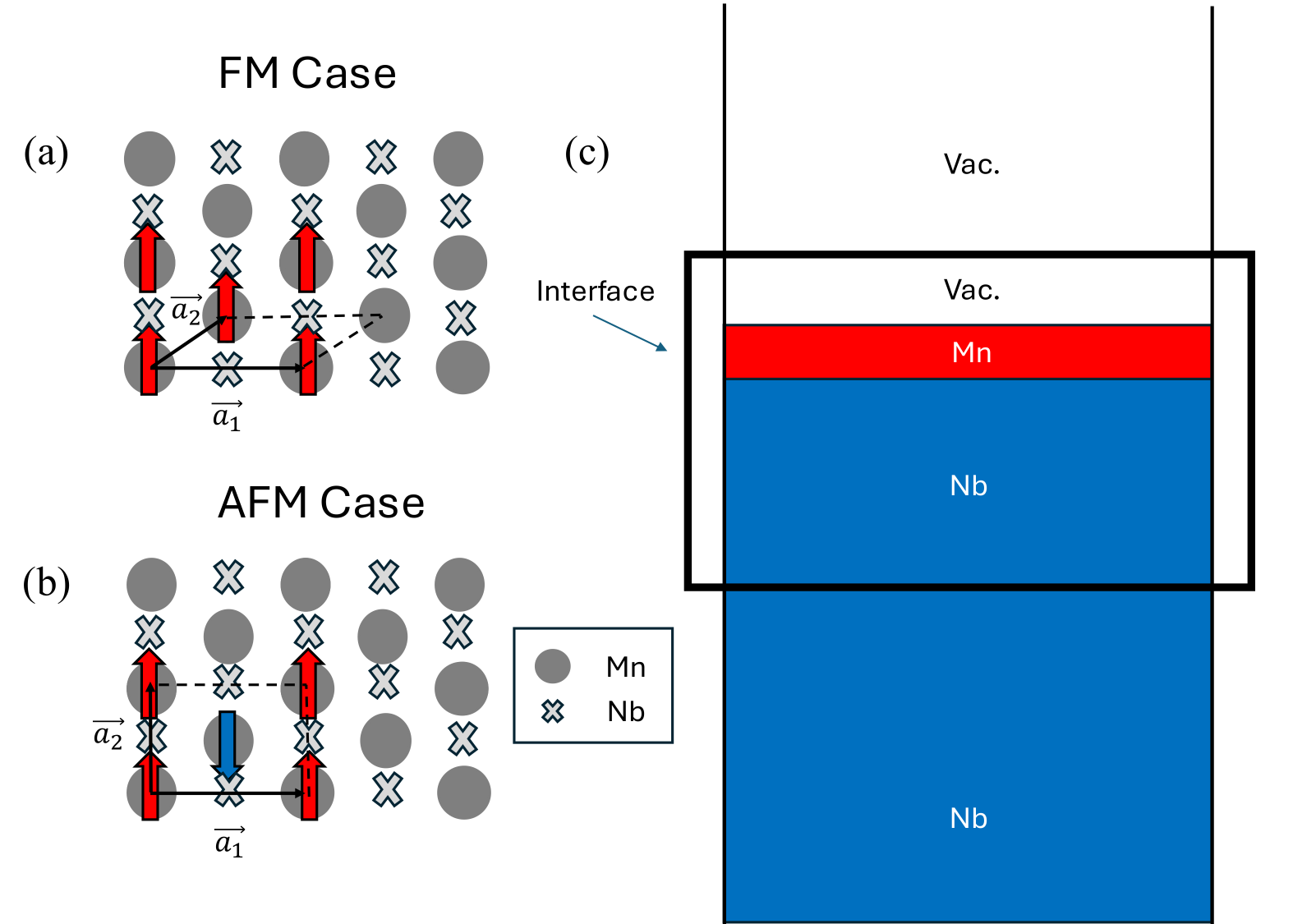}
\caption{(a) A single Mn atomic layer with FM ordering over a Nb(110) substrate. The two-dimensional unit cell is centered-rectangular with $\vec{a_1}=4.6675$ \r{A} along the $[\overline{1}10]$ direction and $\vec{a_2}=2.86$ \r{A} along the $[\overline{1}11]$ direction. The arrows represent the Mn spin moments and are pointing into the page (Red). (b) A single Mn atomic layer with AFM c(2$\times$2) ordering on a Nb(110) substrate. The two-dimensional unit cell is primitive-rectangular with $\vec{a_1}=4.6675$ \r{A} along the $[\overline{1}10]$ direction and $\vec{a_2}=3.3004$ \r{A} along the $[001]$ direction. The arrows represent the Mn spin moments and are pointing into the page (Red) and out of the page (Blue). (c) A schematic diagram of Mn layers on a Nb(110) substrate with the middle interface region consisting of (1-2) Mn layers, some (10-12) Nb layers, as well as several vacuum layers.}
\label{fig:1}
\end{figure}

To address the problem on a microscopic level, we use the SKKR method to solve the (Kohn-Sham) Dirac-Bogoliubov-de Gennes equations within the Oliveira-Gross-Kohn (OGK) superconducting DFT \cite{OGK} for the entire heterostructure. As it is customary in such calculations, the space is divided into three regions: a semi-infinite bulk Nb, an interface region which contains some (10-12) Nb layers to account for the change of the electrostatic potential due to the interface between Nb and Mn layers, the Mn layers themselves (1 or 2), 3 empty spheres layers and finally a semi-infinite vacuum region as shown in Fig.\ref{fig:1}(c). 

In experimental samples of the Mn-Nb(110) heterostructures, a pseudomorphic growth of the Mn layers has been demonstrated \cite{Lo_Conte_2022}, which allows us to use the same two-dimensional lattice constants for the Mn and Nb layers. The structure has a body-centered cubic bcc(110) symmetry with in-plane lattice constants $a=3.3004$ \r{A} along the [001] direction and $b=4.6675$ \r{A} along the $[\overline{1}10]$ direction, and an inter-layer distance of $h=2.337$ \r{A}.

\begin{figure*}
\centering
\includegraphics[width= 0.9 \textwidth]{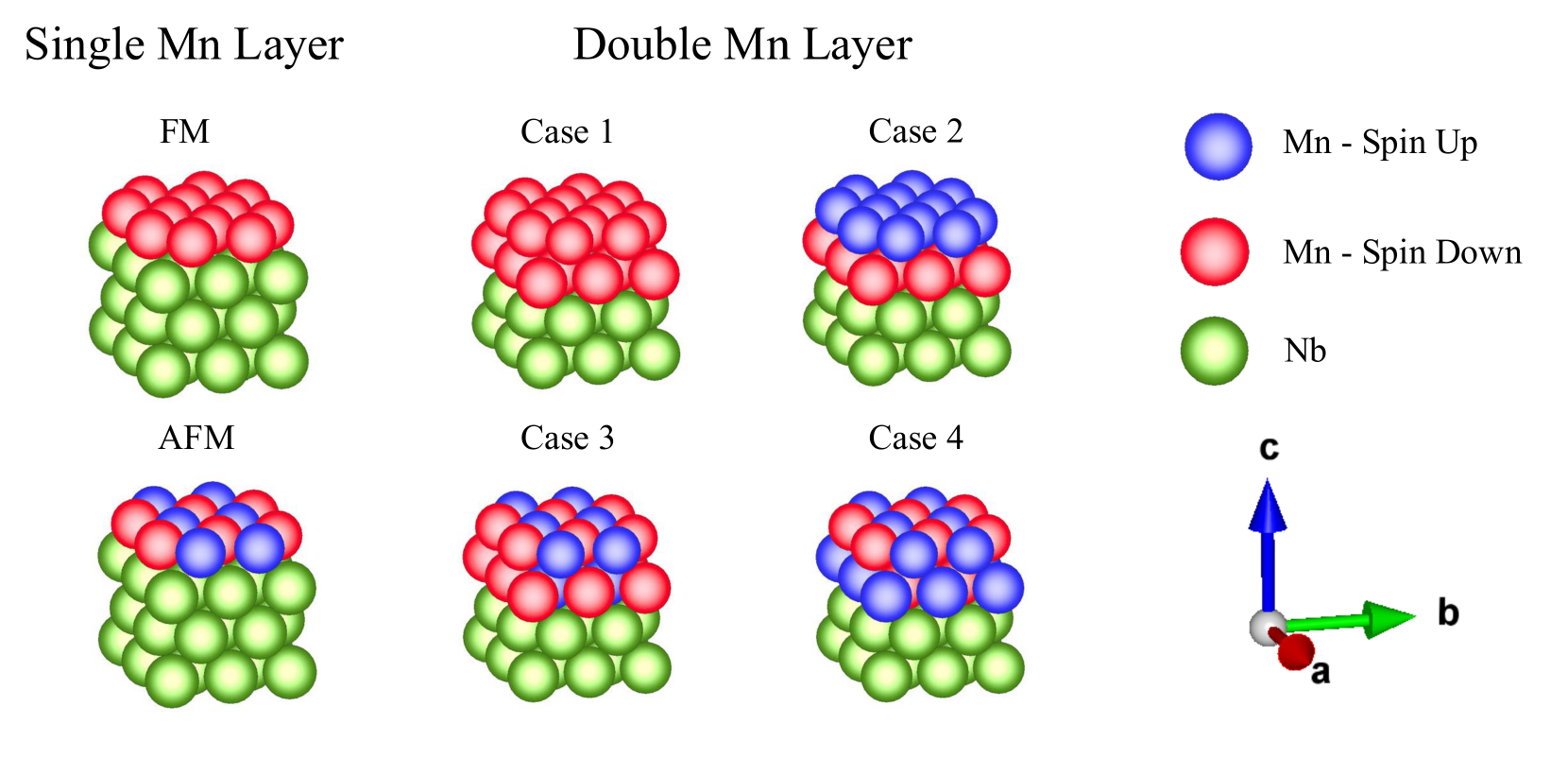}
\caption{
Magnetic ordering cases of the Mn layer(s) in the Mn-Nb heterostructure. The first column shows the single Mn layer cases (FM - AFM), while the second and third columns show the double Mn layer cases. The second column shows the cases with interlayer FM ordering, while the third row shows the cases with interlayer AFM ordering. The first row shows all cases with intralayer FM ordering, while the second row shows all cases with intralayer AFM ordering. }
\label{fig:2}
\end{figure*}

For the single Mn atomic layer heterostructure, we consider two cases: the Mn layer has FM ordering (Fig. \ref{fig:1}(a)) or AFM c(2$\times$2) ordering (Fig. \ref{fig:1}(b)). The latter case matches the experimental data \cite{Lo_Conte_2022} where spin-up and spin-down FM rows along the [001] direction alternate with each other along the $[\overline{1}10]$ direction. For the double Mn layer heterostructure, we consider four cases: the Mn layers have intra- and interlayer FM ordering (in short, Case 1), intralayer FM and interlayer AFM ordering (Case 2), intralayer AFM c(2$\times$2) and interlayer FM ordering (Case 3), and finally, intralayer AFM c(2$\times$2) and interlayer AFM ordering (Case 4). Here Case 4 matches the experimental findings in Ref.~\onlinecite{Lo_Conte_2022}. All the magnetic cases of interest are schematically shown in Fig. \ref{fig:2}. In the intralayer FM cases, we choose a centered-rectangular two-dimensional unit cell with one atom per layer, whereas, for the intralayer AFM c(2$\times$2) case, we choose a primitive-rectangular unit cell with two atoms per layer to correctly describe the AFM pairing. The unit cells and corresponding lattice vectors are shown in Fig.\ref{fig:1}(a) and (b).

In the practical calculations, first, the potentials are converged in the normal state. In this case, we use the Ceperley-Alder exchange-correlation functional \cite{Ceperley1980} as parametrized by Perdew and Zunger \cite{Perdew1981} with an orbital angular momentum cutoff $l_{max}=2$, and the radii for the Nb and Mn atoms $=$ 1.6250~\AA~. A regular $k$-mesh was used corresponding to the different two-dimensional unit cells, with 666 $k$-points in the two-dimensional irreducible Brillouin zone for the intralayer FM case and 1225 $k$-points for the intralayer AFM case. It should be noted that induced moments are observed on the Nb layers. The converged potentials are then used in the SC state calculations to solve the Bogoliubov-de Gennes equations with a step-function pairing potential where bulk pairing potential is used for all Nb layers (experimental value $\Delta_{bulk}=0.112 \text{ mRy or} 1.52 \text{ meV}$ \cite{DeltaBulkforNb}) and 0 for Mn and vacuum layers. One self-consistent field run is completed in the SC state to compute the SC order parameter. Note that the SC pairing potential, the SC order parameter, and the SC gap, are not equivalent quantities for inhomogeneous superconductors, although they are equivalent for homogeneous superconductors. The normal state DOS is calculated using a broadening of 1 mRy (i.e. the imaginary part of the energy), while the quasiparticle DOS in the SC state is calculated using a broadening of 10 $\mu$Ry.

\section{Results}

In this section, we present our results for the systems in the normal state, including the DOS and the magnetic moment for each layer. Then, we move on to show the superconducting state results, including the Bloch Spectral Function (BSF), the SC quasiparticle DOS, and finally, the calculated SC order parameters.

\subsection{Normal-State Properties}

\begin{figure*}
\centering
\includegraphics[width= 0.9 \textwidth]{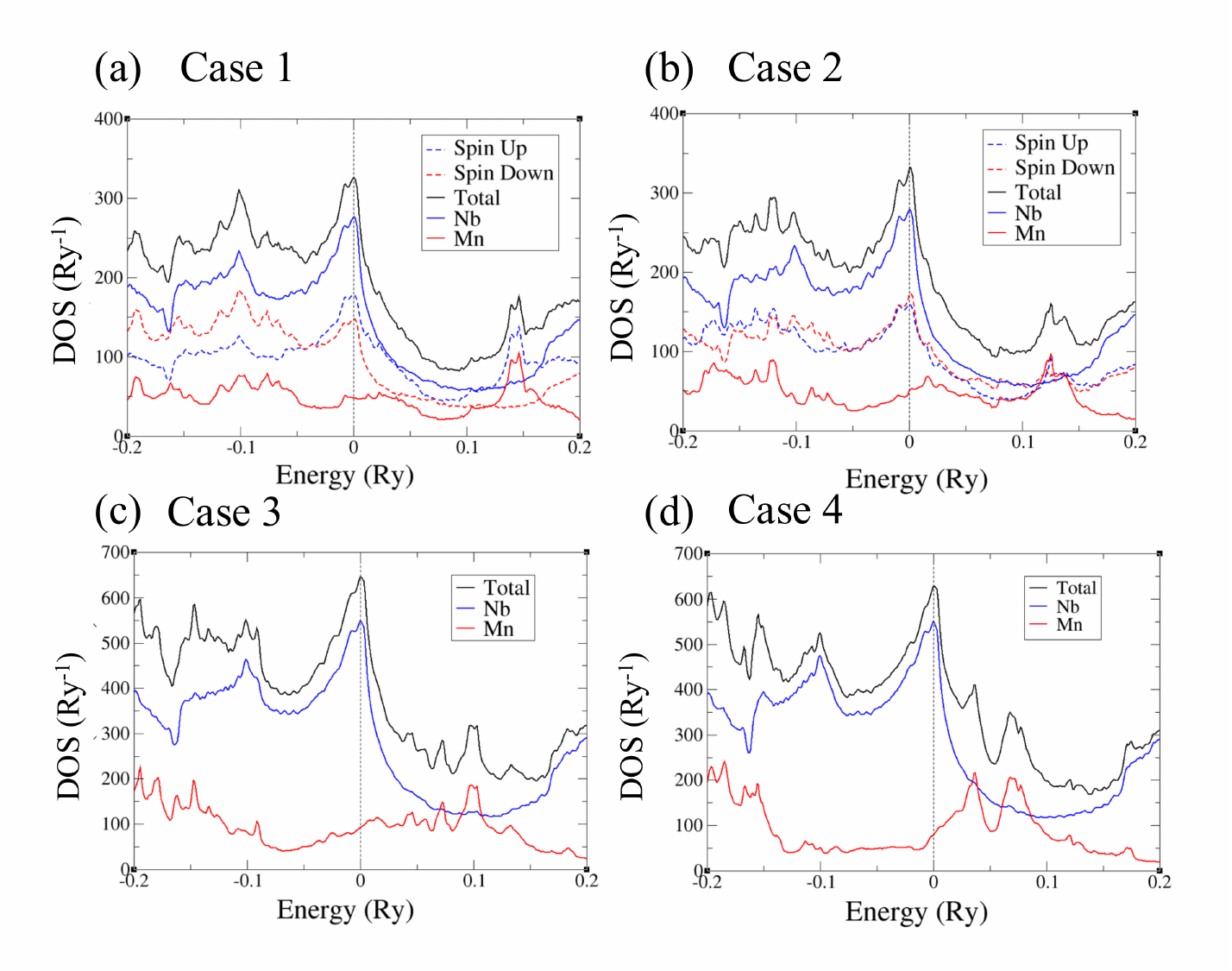}
\caption{Normal-state DOS for the double-Mn-layer heterostructure showing element contributions (sum of the DOS for all layers of the same element in the interface region; 11 Nb layers, and 2 Mn layers). Additionally, spin contributions are indicated for the cases with intralayer ferromagnetic coupling.}
\label{fig:3}
\end{figure*}

First, the normal-state DOS is calculated for the single- and double-Mn-layer heterostructures. Here, we show only the DOS for the double layer cases in Fig.~\ref{fig:3} due to the similarities with the single layer DOS in the cases with the same intralayer magnetic ordering. For the double-Mn-layer heterostructure, the interface region is composed of 11 Nb layers, 2 Mn layers, and 3 vacuum layers, and the DOS shown is the sum of all the interface layers' DOS. For the intralayer FM cases (cases 1 and 2), the exchange splitting in addition to the elemental contributions is shown in the DOS (Fig.~\ref{fig:3}(a) and (b) respectively). Figure~\ref{fig:3}(c) and (d) show the DOS for the intralayer AFM cases (the spin decomposition is not shown as it does not appear in the sum).

\begin{figure*}
\centering
\includegraphics[width= 0.9 \textwidth]{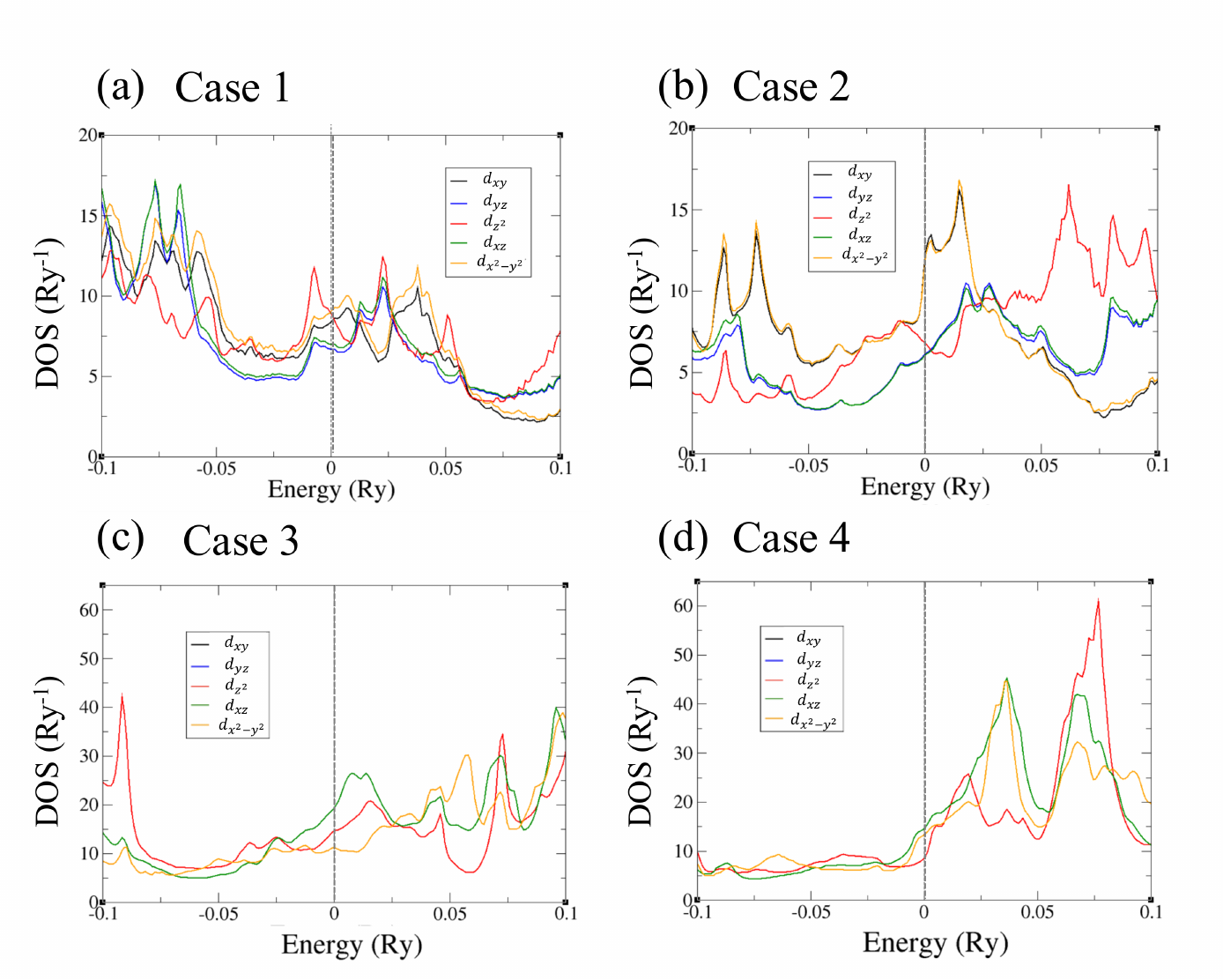}
\caption{DOS projected onto the Mn $3d$-orbitals for the four magnetic cases of the double-Mn-layer heterostructure. }
\label{fig:4}
\end{figure*}

We also calculate the DOS projected onto the Mn $3d$ orbitals for both heterostructures. For the Mn monolayer cases, the $d_{z^2}$ orbital is dominant at the Fermi level, for both the FM and AFM ordering. Figure~\ref{fig:4} shows the DOS projected onto the Mn $3d$ orbitals for all the double-Mn-layer cases.  For Case 1, the $d_{z^2}$ and $d_{x^2-y^2}$ orbitals are dominant at the Fermi level, while for Case 2, the $d_{xy}$ and $d_{x^2-y^2}$ orbitals are dominant. For Cases 3 and 4, $d_{xz}$ and $d_{yz}$ are dominant at the Fermi level.

Table~\ref{table1} lists the calculated magnetic moments of the Mn layers and those induced in the two topmost Nb layers for the Mn single-layer and double-layer heterostructures. Beyond that, the induced moments are indeed negligible. 

For the FM single-layer case, the Mn magnetic moment is about -3.7 $\mu_{\rm B}$ within the Wigner-Seitz radius, and the topmost Nb and next Nb layers have induced magnetic moments of 0.25 $\mu_{\rm B}$ and -0.05 $\mu_{\rm B}$, respectively. For the AFM single-layer case, the magnetic moments of the Mn atoms and the topmost Nb have similar magnitudes to those in the FM case, while the next Nb layer drops to about half the magnitude of the corresponding layer in the FM case, and keeps the direction of the topmost Nb layer's induced moments opposite to the Mn magnetic moment. 

For the double-Mn-layer cases, the topmost Mn atom has a magnetic moment of 3.9-4.0 $\mu_{\rm B}$ in magnitude, while the bottommost Mn atom has a moment of 3.3-3.5 $\mu_{\rm B}$ in magnitude. The induced magnetic moments of the topmost Nb and next Nb layers or atoms show similar patterns to those in the monolayer cases with regards to direction, and similar magnitudes albeit slightly less. The induced moments deeper into the Nb layers show oscillatory behaviour but are extremely small in all systems considered.  

\FloatBarrier

\begin{table}
\centering
 \caption{Magnetic moments of the Mn layer/s and the two topmost Nb layers in units of Bohr magneton ($\mu_B$), where Nb\#2 is the topmost Nb layer and Nb\#1 is the one directly below it, Mn\#1 is the bottommost/first Mn layer directly above the Nb substrate and Mn\#2 is the topmost Mn layer. For the intralayer AFM cases: AFM single-layer, case 3, and case 4, two values are reported for each layer since there are two atoms per layer in the unit cell. \\}
 \begin{tabular}{|c|c|c|c|c|}
 \hline
 Magnetic Case & Nb\#1 & Nb\#2 & Mn\#1 & Mn\#2 \\ [1ex]
 \hline\hline
 FM & -0.049 & 0.249 & -3.708 & - \\
 \hline
 AFM & \begin{tabular}{@{}c@{}} 0.022 \\ -0.022\end{tabular} & \begin{tabular}{@{}c@{}} 0.265 \\ -0.265 \end{tabular} & \begin{tabular}{@{}c@{}} -3.699 \\ 3.699 \end{tabular} & - \\
 \hline
 Case 1 & -0.041 & 0.199 & -3.425 & -4.032 \\
 \hline
 Case 2 & -0.031 & 0.186 & -3.450 & 3.939 \\ [1ex]
 \hline
 Case 3 & \begin{tabular}{@{}c@{}} 0.016 \\ -0.016 \end{tabular} &  \begin{tabular}{@{}c@{}} 0.208 \\ -0.208 \end{tabular} & \begin{tabular}{@{}c@{}} -3.344 \\ 3.344 \end{tabular} & \begin{tabular}{@{}c@{}} -3.944 \\ 3.944 \end{tabular}  \\ [1ex]
 \hline
 Case 4 & \begin{tabular}{@{}c@{}} -0.022 \\ 0.022 \end{tabular} & \begin{tabular}{@{}c@{}} -0.244 \\ 0.244 \end{tabular} & \begin{tabular}{@{}c@{}} 3.462 \\ -3.462 \end{tabular} & \begin{tabular}{@{}c@{}} -4.018 \\ 4.018 \end{tabular} \\ [1ex]
 \hline
\end{tabular}
\label{table1}
\end{table}

\subsection{BSF for the Mn monolayer heterostructures in the SC state}

The Bloch spectral function can be calculated directly from the Green function, and is defined as $A_B(\epsilon, \vec k)= \sum_n \delta(\epsilon -\epsilon_n (\vec k))$. In a layered system, it becomes layer dependent \cite{Csire_2015} while $\vec k$ is in the 2D Brillouin zone: 
\begin{equation}
  A_B^I(\epsilon, \vec k_{||})=
    -\frac{1}{\pi}
    \mathrm{Im}~\mathrm{Tr} \int \dd^{~ \! 3}\! r ~ G^{+}_{II}(\epsilon, \vec r,\vec k_{||}),
\label{eq:BSF}
\end{equation}

where $I$ is the layer index.
Since the BSF is equivalent to the quasiparticle spectrum, a contour plot of the BSF as a function of energy along specified directions of a two-dimensional Brillouin zone is a powerful tool to visualize the quasiparticle states. In a layered system, this can be done for each layer and then optionally summed to get the total BSF, which is what we do here. The spectral functions were calculated by adding a small imaginary part of $\sim$ 0.000005~Ry to the energy.

\begin{figure*}
\centering
\includegraphics[width= 0.9 \textwidth]{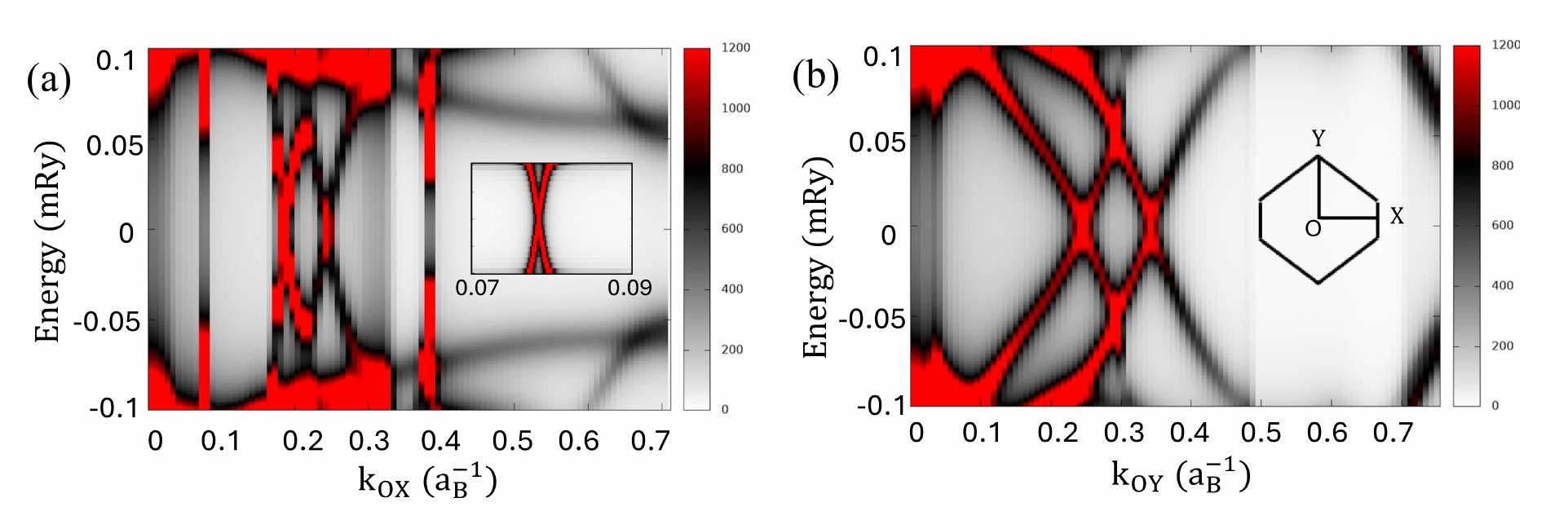}
\caption{BSF for the Mn monolayer heterostructure in the SC state - FM case (a) along the $x$-direction in $k$-space in units of inverse Bohr radius or $a_B^{-1}$, (b) along the $y$-direction in $k$-space, within the energy range of the bulk Nb SC gap. The inset in (a) shows a zoom-in for 0.07 $a_B^{-1}$ $< k_{\text{OX}} <$ 0.09 $a_B^{-1}$, while the inset in (b) shows the 2D irreducible BZ in the FM case~\cite{Cunningham_1974}.}
\label{fig:5}
\end{figure*}

Figure~\ref{fig:5}(a) and (b) show the total calculated BSF in the SC state within the bulk Nb SC gap along the $k_{\text{OX}}$ and $k_{\text{OY}}$ directions in the 2D Brillouin zone for the single Mn layer heterostructure in the FM case. For both directions and all layers (Nb and Mn), there are Andreev bound states consisting of both electron- and hole-like components within the SC gap. Additionally, there are bound states that even cross the Fermi level at $k_{\text{OX}} \approx 0.08 \ a_B^{-1}, 0.19 \ a_B^{-1}, 0.24 \ a_B^{-1}$, and at $k_{\text{OY}} \approx 0.24 \ a_B^{-1}, 0.34 \ a_B^{-1}$. However, the band structure around $k_{\text{OX}} \approx 0.39 \ a_B^{-1}$ can be mistaken for a crossing, but after carefully reviewing the raw data close to the Fermi level, it was determined that it is not a real crossing; there is a relatively small gap opening up. Interestingly, for the Mn layer, both electron and hole components of the BSF are found along both the $k_{\text{OX}}$ and the $k_{\text{OY}}$ directions, but the spin contributions to the Fermi level crossing states were found to be unequal, with one spin component heavily contributing to the crossing around $k_{\text{OX}} = 0.08 \ a_B^{-1}$ and the other heavily contributing to the rest of the crossings at $0.15 \ a_B^{-1} < k_{\text{OX}} < 0.3 \ a_B^{-1}$ and $0.05 \ a_B^{-1} < k_{\text{OY}} < 0.5 \ a_B^{-1}$. For the Nb layers, a similar trend was seen for the electron and hole contributions. However, for the spin contributions, the $k_{\text{OX}}$ direction had similar trends, while the $k_{\text{OY}}$ direction saw equal spin contributions overall. Such in-gap features can be thought of as Shiba bands \cite{Yu, Shiba, Rusinov}. Further analysis and BSF figures are presented in section II of the supplementary material for the interested reader.  

\begin{figure*}
\centering
\includegraphics[width= 0.9 \textwidth]{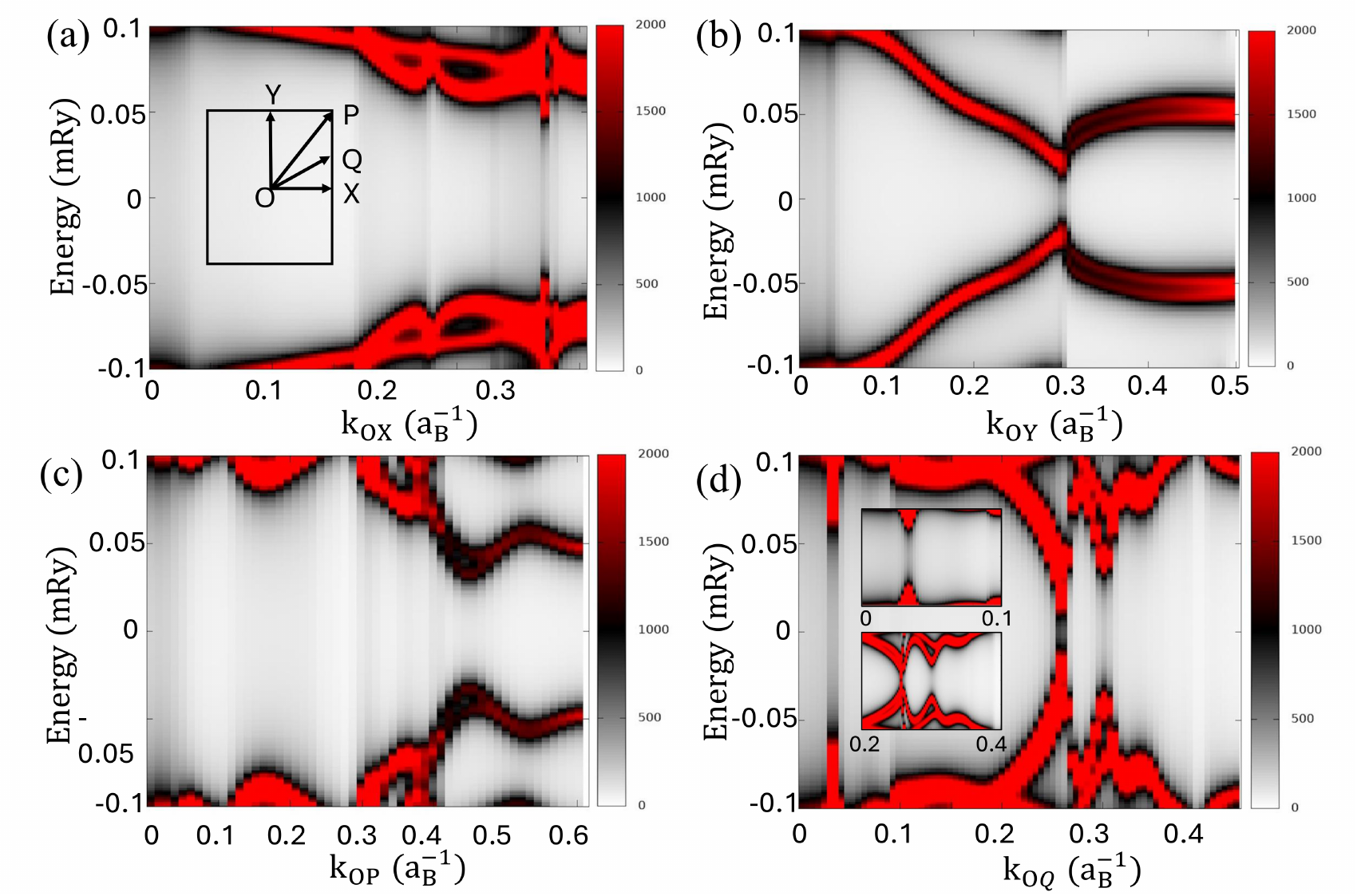}
\caption{BSF for the Mn monolayer heterostructure in the SC state - AFM case (a) along the $x$-direction in $k$-space in units of $a_B^{-1}$, (b) along the $y$-direction, (c) along the OP-direction, and (d) along the OQ-direction [Q=(0.356, 0.252)$a_B^{-1}$], within the bulk Nb SC gap. The inset in (a) shows the 2D irreducible BZ in the AFM case~\cite{Cunningham_1974}.}
\label{fig:6}
\end{figure*}

Figure~\ref{fig:6}(a)-(d) show the calculated BSF in the SC state along the $k_{\text{OX}}$, $k_{\text{OY}}$, $k_{\text{OP}}$, and $k_{\text{OQ}}$ directions for the AFM Mn single layer heterostructure, respectively, where $k_{\text{OP}}$ and $k_{\text{OQ}}$ are the directions illustrated in the inset of Fig. \ref{fig:6}(a). For the $k_{\text{OY}}$, $k_{\text{OX}}$, $k_{\text{OP}}$ directions, there are Andreev bound states with an induced SC gap within the bulk Nb gap. The induced SC gap size depends on the direction in $k$-space. Interestingly, there is a Fermi level crossing along the OQ direction at $k_{\text{OQ}} = 0.26 \ a_B^{-1}$, which is unexpected considering the AFM nature of the heterostructure. Again, both electron and hole components of the BSF are found along all directions and layers just as expected from Andreev scattering. See section II of the supplementary material for further analysis and more BSF figures.

\subsection{SC quasiparticle DOS}

\begin{figure*}
\centering
\includegraphics[width= 0.9 \textwidth]{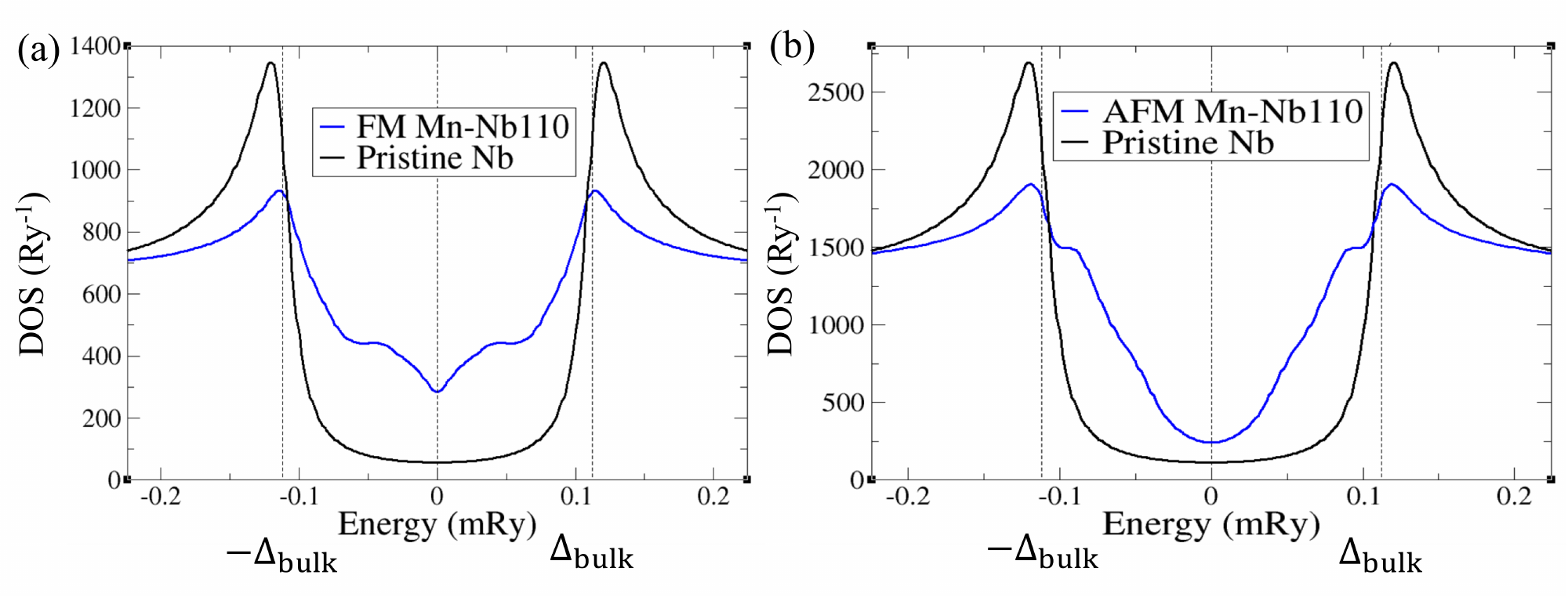}
\caption{ (a) SC DOS for pristine Nb (black) and the single-Mn-layer heterostructure - FM case (blue). (b) SC DOS for pristine Nb (black) and the single-Mn-layer heterostructure - AFM case (blue).}
\label{fig:7}
\end{figure*}

\begin{figure*}
\centering
\includegraphics[width= 0.9 \textwidth]{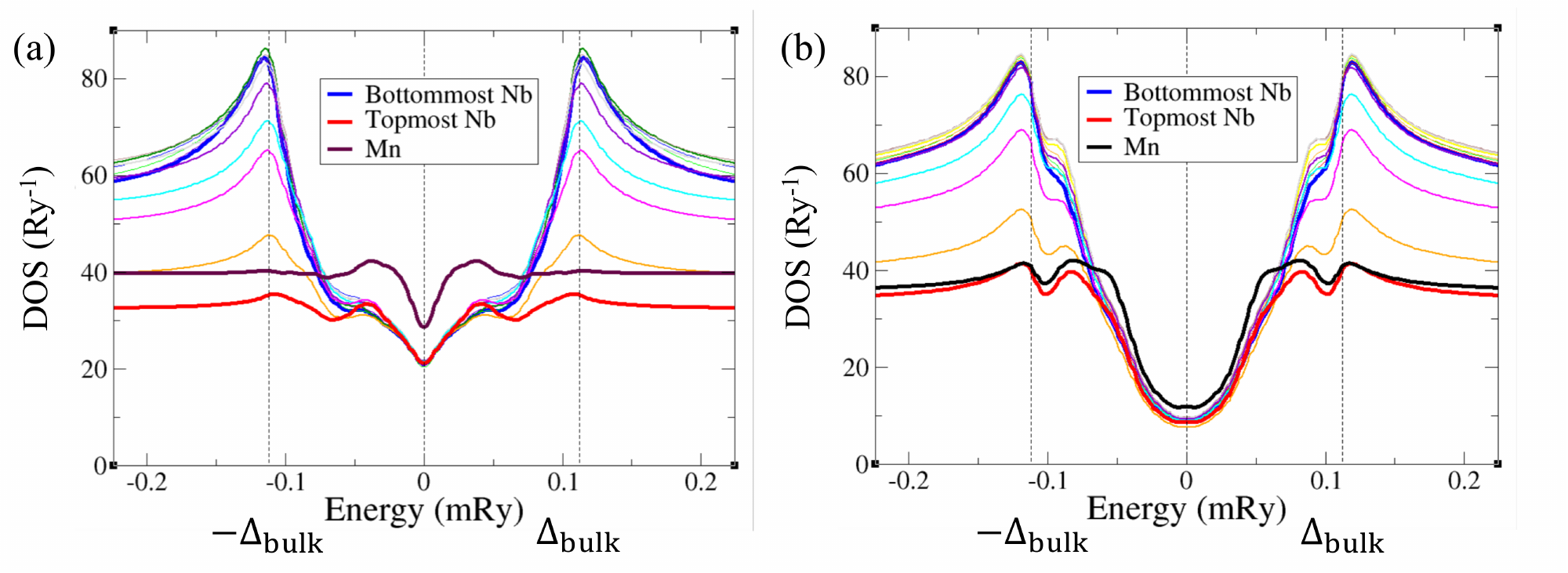}
\caption{ (a) Layer-dependent DOS for the FM case single-Mn-layer heterostructure. (b) Layer-dependent DOS for the AFM case single-Mn-layer heterostructure. The bottommost and topmost Nb layers and the Mn layer are highlighted with thicker lines.}
\label{fig:8}
\end{figure*}

\begin{figure*}
\centering
\includegraphics[width= 0.9 \textwidth]{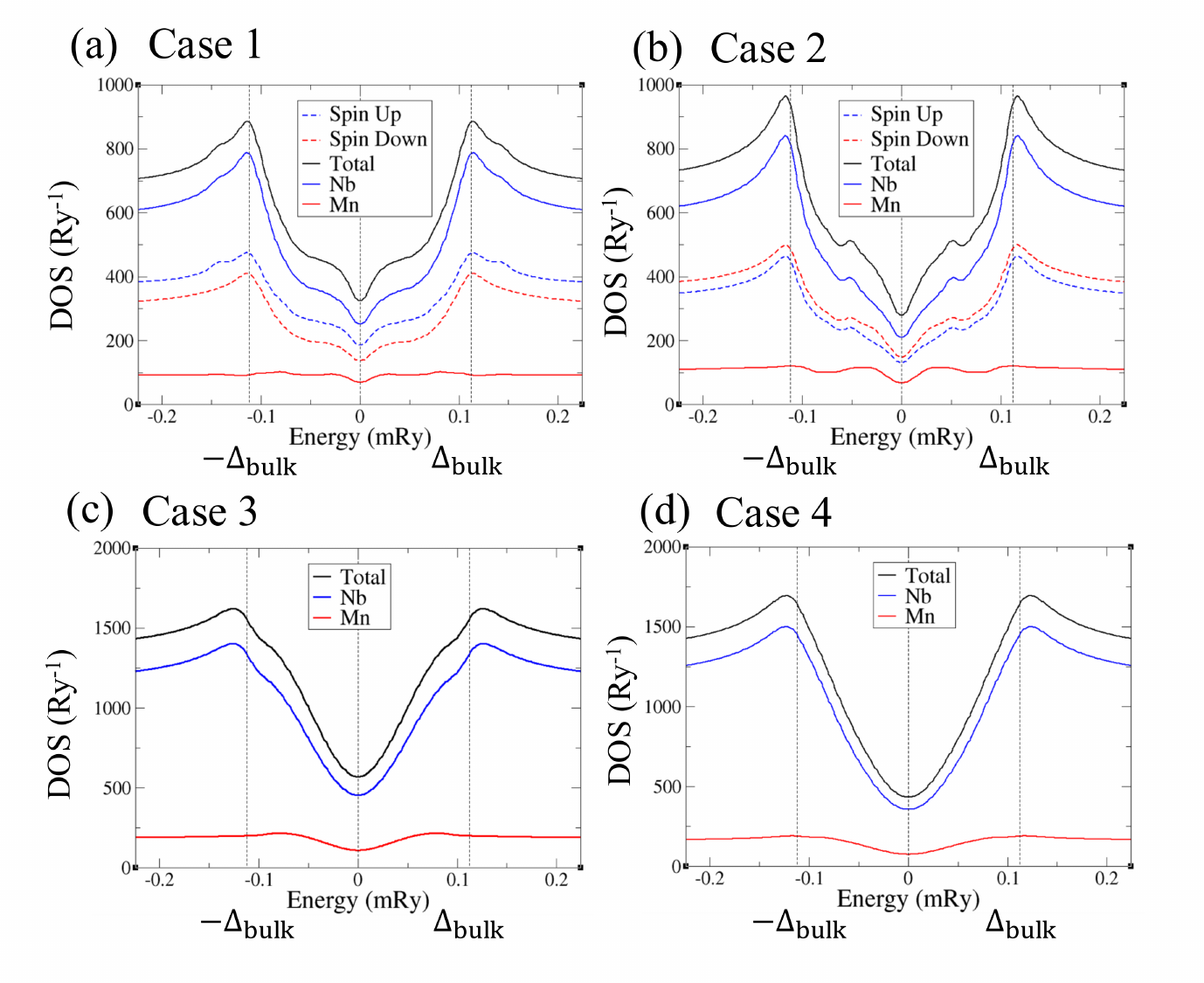}
\caption{SC DOS for the double-Mn-layer heterostructure showing element and spin contributions where relevant in all four magnetic cases}
\label{fig:9}
\end{figure*}

Figure~\ref{fig:7}(a) shows the calculated quasiparticle DOS in the SC state for the FM single-Mn-layer heterostructure in comparison to the quasiparticle DOS of pristine Nb. We observe in-gap states within the bulk Nb SC gap. A flat plateau-like region appears around $\pm$0.06 mRy within the SC gap, and it goes into a deep, sharp V-shape as the energy approaches the Fermi level (taken to be 0), which is similar to the in-gap features found for a Gd-Nb heterostructure in Ref.~\cite{Gd_Park2024}. 

Figure~\ref{fig:7}(b) shows the calculated quasiparticle DOS in the SC state for the AFM single-Mn-layer heterostructure. There is a secondary gap at about $\pm$0.09 mRy within the bulk SC gap. In contrast to the FM case, the in-gap states are not sharp V-shaped but more U-shaped. We note here that these DOS plots are the sum of all the layers' DOS, and numerically, we expect to find close to zero, but non-zero, values for the DOS even in the pristine Nb SC gap. Adding these small effects up gives us the shift in the DOS from zero to a higher value within the SC gap.

To further analyze the SC-state DOS, we show the layer-dependent DOS in Fig.\ref{fig:8} where we highlight the bottommost and topmost Nb layers, and the Mn layer. In the FM case, the peaks near $\pm\Delta_{\rm bulk}$ appear for all Nb layers, whereas the flat plateau-like region occurs for all Nb layers except for the topmost Nb layer interfaced with the Mn layer. The sharp V-shaped in-gap states are also found for all Nb layers and the FM Mn layer. In the AFM case, peaks near $\pm\Delta_{\rm bulk}$ and the secondary gap appear for all Nb layers. The U-shaped in-gap states are also found for all Nb layers and the AFM Mn layer. The DOS of the Mn layer is very close to the DOS of the topmost Nb layer.

Next, we investigate the quasiparticle DOS of the double-Mn-layer heterostructure in the SC state. Figure \ref{fig:9} shows element-specific and spin-up and spin-down contributions to the quasiparticle DOS where relevant as well as the total DOS in Cases 1-4. For Case 1 with the intralayer and interlayer FM ordering, a wider plateau region and a sharper V-shaped dip at the Fermi level, than the FM monolayer case are seen. For Case 2, with intralayer FM ordering and interlayer AFM ordering, a slightly deeper V-dip occurs, accompanied by pronounced peaks at its edges. For Case 3, with the intralayer AFM ordering and interlayer FM ordering, a slight hint of the secondary gap is seen. For Case 4, with the intralayer and interlayer AFM ordering, the secondary gap almost disappears.

When our calculations are compared to the experimental local DOS, dI/dU, obtained using a Cr STM tip \cite{Lo_Conte_2022}, we find a qualitative agreement. The experimental DOS has U-shaped in-gap states, which is similar to our results for the AFM monolayer case and Case 4 (the cases that have the same magnetic ordering as the experimental heterostructure). Interestingly, the peak height near $\pm\Delta_{\rm bulk}$ for Case 4 is somewhat smaller than that for the AFM monolayer, which is consistent with the experimental finding as well.

\begin{figure*}
\centering
\includegraphics[width= 0.8 \textwidth]{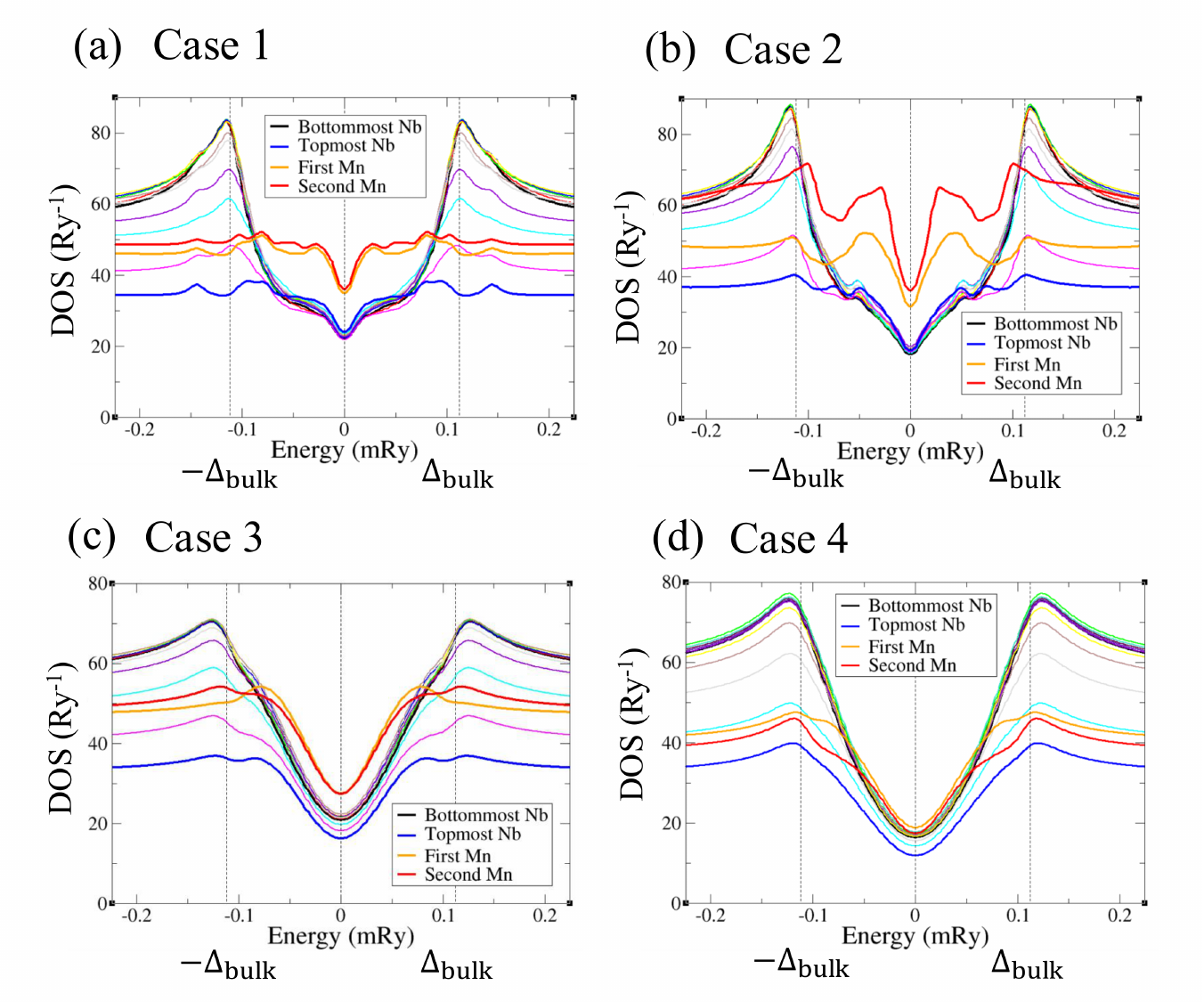}
\caption{Layer-dependent SC-state DOS for the double-Mn-layer heterostructure in all four magnetic cases. The bottommost and topmost Nb layers and the two Mn layers are highlighted with thicker lines.}
\label{fig:10}
\end{figure*}

Furthermore, we may also calculate the layer-dependent DOS, as shown in Fig.\ref{fig:10}, where it is highlighted with thicker lines for the bottommost and topmost Nb layers as well as for the two Mn layers. Similarly to the single layer cases, all the Nb layers bear the features of the in-gap states in all four cases. A clear difference can be seen in the Mn layer DOS in the four cases attributed to the various magnetic states, which ultimately influence the total DOS features. The observed in-gap states are therefore a consequence of the Cooper pairs being affected by the Mn exchange field, and can be thought of as Yu-Shiba-Rusinov (YSR) states \cite{Yu, Shiba, Rusinov} which again agrees with the experimental findings in Ref.~\cite{Lo_Conte_2022}. The YSR states hybridize to form Shiba bands, and that can be seen in the BSF figures. 

\subsection{Superconducting order parameter}

\begin{figure*}
\centering
\includegraphics[width= 0.9 \textwidth]{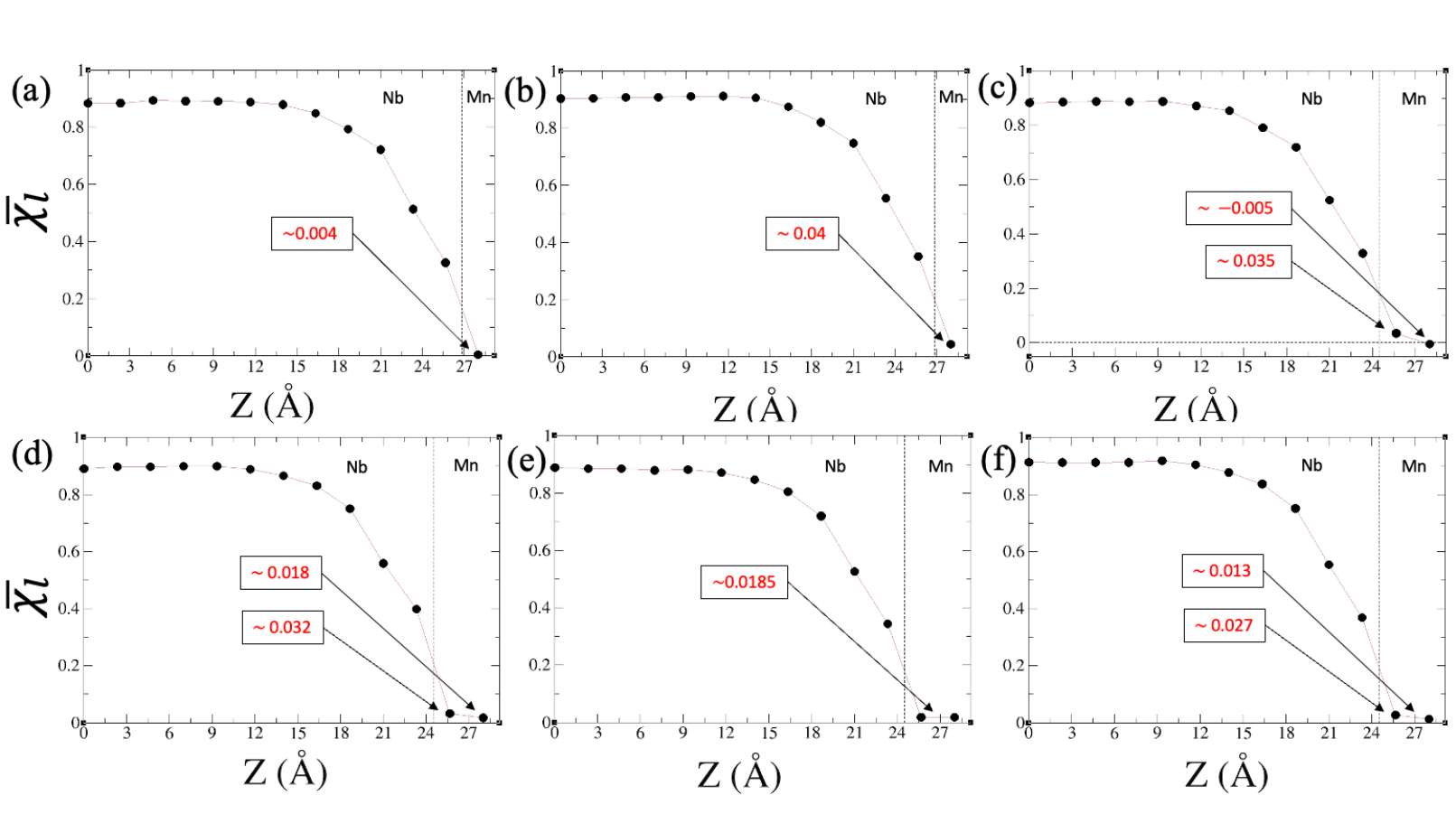}
\caption{Layer by Layer SC order parameter as a function of the vertical $z$-coordinate in the (a) FM case, (b) AFM case for the single Mn layer heterostructure, (c-f) corresponding to Cases 1-4 for the double Mn layer heterostructure.}
\label{fig:11}
\end{figure*}

From the Green function, the layer resolved spin-singlet order parameter (OP) \cite{Csire2018} can be calculated as:
\begin{equation}
    \chi_\text{S}(\varepsilon, n) = - \frac{1}{\pi} \Im~ \mathrm{Tr}_L~\mathcal{S}_s 
    \{ G^{nn,\text{eh}}_{Ls,L's'}(\varepsilon)\} \, 
\end{equation}
where $\mathrm{Tr}_L$ denotes the trace in angular-momentum space, and $\mathcal{S}_s$ generates the spin-singlet,  $\mathcal{S}_s \{f(s,s^\prime)\}=\frac{\sqrt{2}}{2} \left(f(\frac{1}{2},-\frac{1}{2})-f(-\frac{1}{2},\frac{1}{2}) \right)$. Its energy integral up to the Fermi energy is often referred to as the anomalous charge. 

It has already been established that artificially constructed heterostructures do host spin-triplet Cooper pairs \cite{Robinson2010,di_bernardo_signature_2015}. The fermionic nature of the electron dictates that in the case of triplet pairing, the spatial component of the wave function has to be odd. Consequently, in the context of a multi-band Hamiltonian for bulk systems, this allows the possibility of even parity odd orbital triplet pairs, which we may call Internally Antisymmetric Triplet (IAT) states to avoid confusion with the bulk usage of the term, where a full parity operator exists \cite{nyari2023topological, Tanaka_2012}. Such states were shown to be responsible for the experimentally observed simultaneous appearance of magnetism and superconductivity in certain materials \cite{csiretriplet1, csiretriplet2}. 

Therefore, it is expected that the relativistic Andreev scattering process (captured accurately by the generalized multiple scattering theory for the superconducting state) results in IAT states in the Mn layers, which are antisymmetric with regards to the orbital degrees of freedom. This statement is easy to understand because during the solution of the relativistic Bogoliubov de Gennes equation, a mixing occurs between the spin and orbital degrees of freedom together with the electron-hole character. This type of symmetry classification of Cooper pairs is important, since it is necessary to distinguish these features from the odd-frequency spin triplet pairing, which may also appear in many artificial superconductor-magnet hybrid structures as presented in Ref. \cite{Cayao2021}.

Here, we aim to find the dominant component of the 
induced triplet OP in real space, therefore -- mimicking the singlet order parameter -- the following quantity is defined to account for the norm of the energy-resolved IAT order parameter (which is now a matrix in orbital indices):

\begin{equation}
    \overline{\chi}_{\text{IAT} (\varepsilon)} = 
     \sum_n \sum_{i=-1,0,1}
        - \frac{1}{\pi} \  \left| \left| \Im
              \mathcal{A}_L \mathcal{T}_s^i
              \{ G^{nn,\text{eh}}_{Ls,L's'}(\varepsilon)\}
            \right| \right|_\text{F}
\end{equation}

\noindent with the antisymmetrization in angular momentum space,
\begin{equation*}
    \mathcal{A}_L \{ f(L,L^\prime)\}= \{ \frac{1}{2} \left( f(L,L^\prime) -f(L^\prime,L) \right) \}
\end{equation*}
\noindent and the projections on spin-triplets, 
\begin{equation*}
    \mathcal{T}_s^0 \{ f(s,s^\prime)\} =\frac{\sqrt{2}}{2} \left( f(\frac{1}{2},-\frac{1}{2})+f(-\frac{1}{2},\frac{1}{2}) \right)
\end{equation*}
\begin{equation*}
    \mathcal{T}_s^{\pm 1} \{f(s,s^\prime)\} = f(\pm \frac{1}{2}, \pm \frac{1}{2})
\end{equation*}
\noindent while $|| M ||_\text{F}$ denotes the Frobenius norm of matrix $M$. This quantity signals the emergence of IAT pairs in any angular momentum channels and its energy integral gives a quantity analogous to the anomalous charge.

\noindent Accordingly, We first calculate the singlet $\chi_S$ component of the SC order parameter, and then calculate $\overline{\chi^l_S}$ for each layer $l$, by taking a ratio between the SC order parameter calculated for that layer and the corresponding order parameter of the bulk Nb,

\begin{equation}
    \overline{\chi^l_S}=\frac{\int_{WS} \chi_S^l(r) \,dr}{\int_{WS} \chi_S^{Nb}(r) \,dr}
\end{equation}

\noindent where the integral runs over the atomic sphere of the Wigner-Seitz radius. 

Figure \ref{fig:11}(a) and (b) show the calculated singlet component of the SC order parameter for the single Mn layer heterostructure (1 Mn layer and 12 Nb layers) relative to that for bulk Nb for the FM and AFM cases, respectively. The value for the bottommost interface Nb layer is about 90\% of that of the bulk Nb. The value starts decreasing close to the interface until it reaches its lowest at the Mn layer where there is an order of magnitude difference in the value of the singlet component of the SC order parameter between the FM and AFM cases. 

For the double Mn layer heterostructure, we see similar trends in Fig.\ref{fig:11}(c), (d), (e) and (f) with overall slightly smaller values. The most notable difference is in case 1 (Fig.\ref{fig:11}(c)) where there is FM intralayer and interlayer pairing); the SC order parameter takes a negative value at the second Mn layer which seems to suggest that there will be oscillations in the order parameter as we increase the FM paired Mn layers.

Next, we calculate the IAT component and take the ratio between the triplet component and the singlet component in the Mn layers for each case,

\begin{equation}    \overline{\chi^{Mn}_{0,\pm1}}=\frac{\chi^{Mn}_{0,\pm1}}{\chi_S^{Mn}} 
\end{equation}

Table \ref{table2} shows the calculated ratios and their sum ($\overline{\chi_{\text{IAT}}^{Mn}}$). We notice that $\overline{\chi_{\text{IAT}}^{Mn}}$ frequently is greater than 1, and is never less than $\sim$ 0.7, which indicates a singlet-triplet mixing in the Mn layers of the heterostructure with the two highest values corresponding to the FM single-layer case and case 1 (double-layer case with intralayer and interlayer FM ordering). However, note that the values are still very small, as the singlet component in the Mn layer/s only reaches a maximum of 4.44\% of the bulk Nb singlet component value as seen in Fig.~\ref{fig:11}(b). For more details, see section III in the supplementary material.

\section{Conclusion}

\begin{table}
 \caption{Triplet to Singlet component ratios for each Mn layer ($\overline{\chi^{Mn}_{0,\pm1}}=\frac{\chi^{Mn}_{0,\pm1}}{\chi_S^{Mn}}$), and the sum of those ratios for all magnetic cases ($\overline{\chi_{\text{IAT}}^{Mn}}$) \\}
 \begin{tabular}{|c|c|c|c|c|}
 \hline
 Magnetic Case & $\overline{\chi_0^{Mn}}$ & $\overline{\chi_1^{Mn}}$ & $\overline{\chi_{-1}^{Mn}}$ & $\overline{\chi_{\text{IAT}}^{Mn}}$\\ [1ex]
 \hline\hline
 FM & 5 & 0.364 & 3.295 & 8.659\\
 \hline
 AFM & 0.605 & 0.083 & 0.083 & 0.771\\
 \hline
 Case 1 & \begin{tabular}{@{}c@{}} 0.294 \\ 0.979 \end{tabular} & \begin{tabular}{@{}c@{}} 0.126 \\ 0.319 \end{tabular} & \begin{tabular}{@{}c@{}} 0.279 \\ 1.979 \end{tabular} & \begin{tabular}{@{}c@{}} 0.699 \\ 3.277 \end{tabular}\\
 \hline
 Case 2 & \begin{tabular}{@{}c@{}} 0.871 \\ 0.488 \end{tabular} & \begin{tabular}{@{}c@{}} 0.139 \\ 0.365 \end{tabular} & \begin{tabular}{@{}c@{}} 0.355 \\ 0.059 \end{tabular} & \begin{tabular}{@{}c@{}} 1.365 \\ 0.912 \end{tabular}\\ [1ex]
 \hline
 Case 3 & \begin{tabular}{@{}c@{}} 0.648 \\ 0.698 \end{tabular} & \begin{tabular}{@{}c@{}} 0.331 \\ 0.220 \end{tabular} & \begin{tabular}{@{}c@{}} 0.331 \\ 0.220 \end{tabular} & \begin{tabular}{@{}c@{}} 1.31 \\ 1.138 \end{tabular}\\ [1ex]
 \hline
 Case 4 & \begin{tabular}{@{}c@{}} 0.692 \\ 0.515 \end{tabular} & \begin{tabular}{@{}c@{}} 0.110 \\ 0.546 \end{tabular} & \begin{tabular}{@{}c@{}} 0.110 \\ 0.546 \end{tabular} & \begin{tabular}{@{}c@{}} 0.912 \\ 1.607 \end{tabular}\\ [1ex]
 \hline
\end{tabular}
\label{table2}
\end{table}

In this paper, we show the influence of different magnetic orders on proximity-induced
superconductivity in Mn single and double layers on top of a Nb(110) substrate by using the recently developed DBdG solver within the SKKR Green's function formalism. First, we showed the normal-state DOS and the Mn-$3d$-orbitals projected DOS for the double-Mn-layer heterostructure, and the differences between each magnetic case. We also found that there are induced magnetic moments in the two nearest Nb layers to the Mn owing to the Mn moments. 

Next, we investigated the SC-state properties; We computed the momentum-resolved Bloch spectral function for the single-Mn-layer heterostructure and found multiple Fermi-level crossing bands in the ferromagnetic Mn layer case and one in the antiferromagnetic Mn case. We then showed the SC-state DOS for both the single- and double-Mn-layers heterostructures, where we found interesting in-gap features associated with intralayer ferromagnetic ordering, namely a plateau and a V-dip at the Fermi level, and the appearance of a secondary gap and U-shaped in-gap states associated with the intralayer AFM ordering. 

Lastly, we saw an induced small finite value for the singlet SC order parameters in the Mn layer/s with the maximum being only 4.44\% of the bulk value, and even found a negative value in the double Mn layer heterostructure with intralayer and interlayer FM ordering (case 1), suggesting the possibility for observing order parameter oscillations with more Mn layers. We also observed some singlet-triplet mixing in the Mn layer/s where the calculated triplet (IAT) to singlet component ratios were found to be significant, with higher triplet (IAT) values associated with the FM ordering, keeping in mind that the absolute values are small compared to the Nb singlet value (two or three orders of magnitudes less). This magnetic-order-dependent -albeit negligible in magnitude- SC order parameter demonstrates that magnetic order can tune the balance between singlet and triplet correlations. 

\section*{Acknowledgement} 

S.E. gratefully acknowledges support from the Department of Physics at Virginia Tech, and support from the Austrian Science Fund (FWF) project No. P33491-N “ReCALL”. The computational support was provided by the Virginia Tech
Advanced Research Computing (ARC) and through allocation DMR060009N from the Advanced Cyberinfrastructure
Coordination Ecosystem: Services and Support (ACCESS)
program \cite{Access}, which is supported by National Science Foundation Grants No. 2138259, No. 2138286, No. 2138307, No. 2137603, and No. 2138296. B.U. acknowledge financial support by the National Research, Development, and Innovation Office (NRDI Office) of Hungary under Project No. K142652.

\bibliography{citations}

\end{document}


\special{header=psfrag.pro}

\title{Supplemental material for Influence of Magnetic Order on Proximity-Induced Superconductivity in
Mn Layers on Nb(110) from First Principles}

\author{Sohair ElMeligy}
\email{sohaire@vt.edu}
\affiliation{Department of Physics, Virginia Tech, Blacksburg, Virginia 24061, USA}
\author{Bal\'azs \'Ujfalussy}
\email{ujfalussy.balazs@wigner.hu}
\affiliation{Wigner Research Centre for Physics, Institute for Solid State Physics and Optics, H-1525 Budapest, Hungary}
\author{Kyungwha Park}
\email{kyungwha@vt.edu}
\affiliation{Department of Physics, Virginia Tech, Blacksburg, Virginia 24061, USA}

\maketitle

\section{Normal State DOS}

\begin{figure*}
\centering
\includegraphics[width= 0.9 \textwidth]{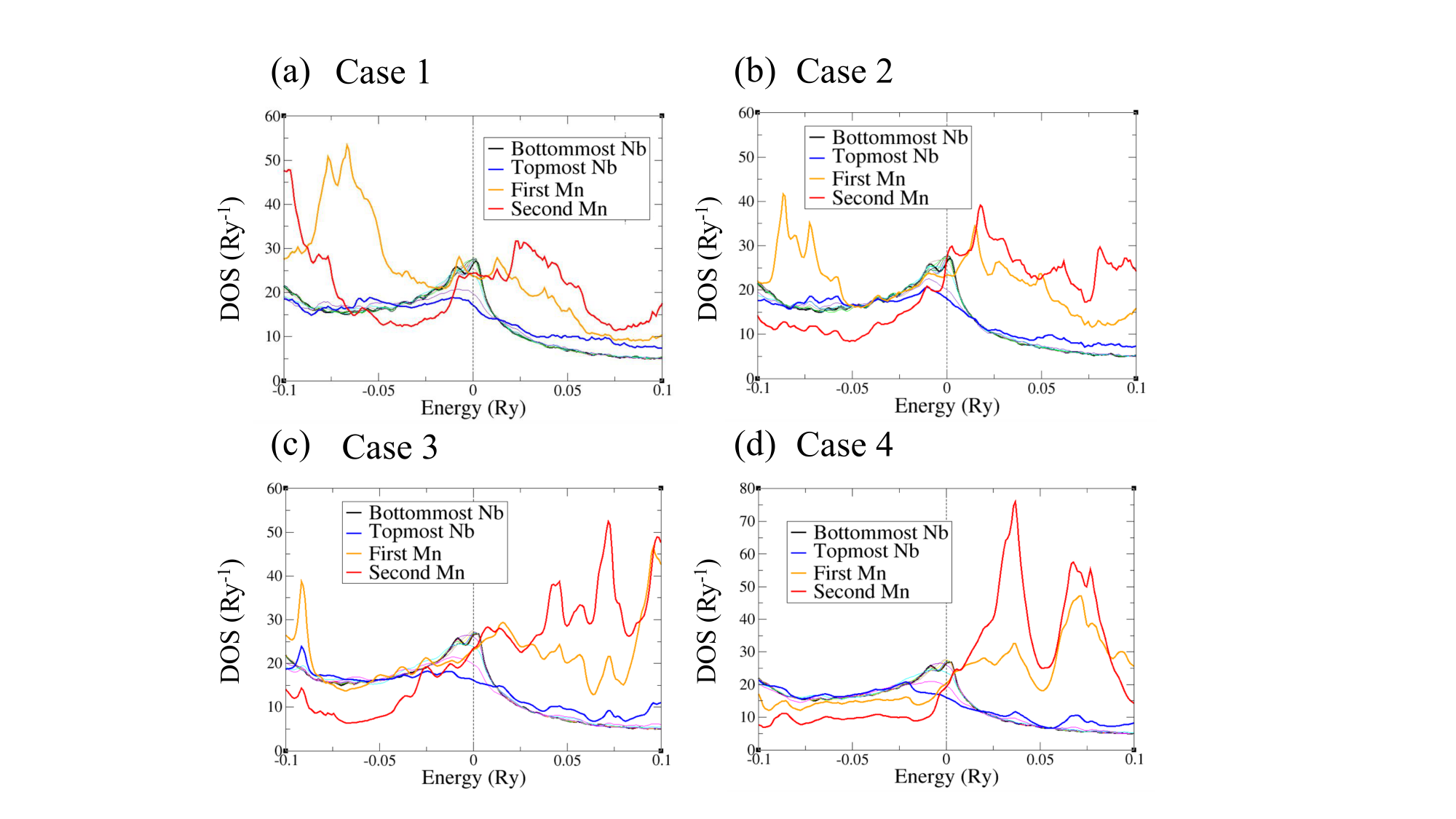}
\caption{Normal-state layer-dependent DOS for the double-Mn-layer heterostructure for (a) case 1 - intralayer and interlayer FM ordering, (b) case 2 - intralayer FM and interlayer AFM ordering, (c) case 3 - intralayer AFM and interlayer FM ordering, and (d) case 4 - intralayer and interlayer AFM ordering. The bottommost and topmost Nb layers, and the Mn layers are highlighted with thicker lines.
}
\label{fig:17}
\end{figure*}

Additional figures and analysis for the normal state DOS are presented in this section; A layer resolved DOS for all four magnetic cases in the double-Mn-layer heterostructure is shown in Fig.\ref{fig:17} with the following layers highlighted with thicker lines: Bottommost and topmost Nb layers, First and second Mn layers. We note that the contribution to the DOS from the bottommost Nb layer is mostly unchanged for all cases, whereas the topmost Nb layer is more affected by the magnetic order of the Mn layers, with the DOS at the Fermi level $E_F$ remaining more or less the same throughout. 

\begin{figure*}
\centering
\includegraphics[width= 0.9 \textwidth]{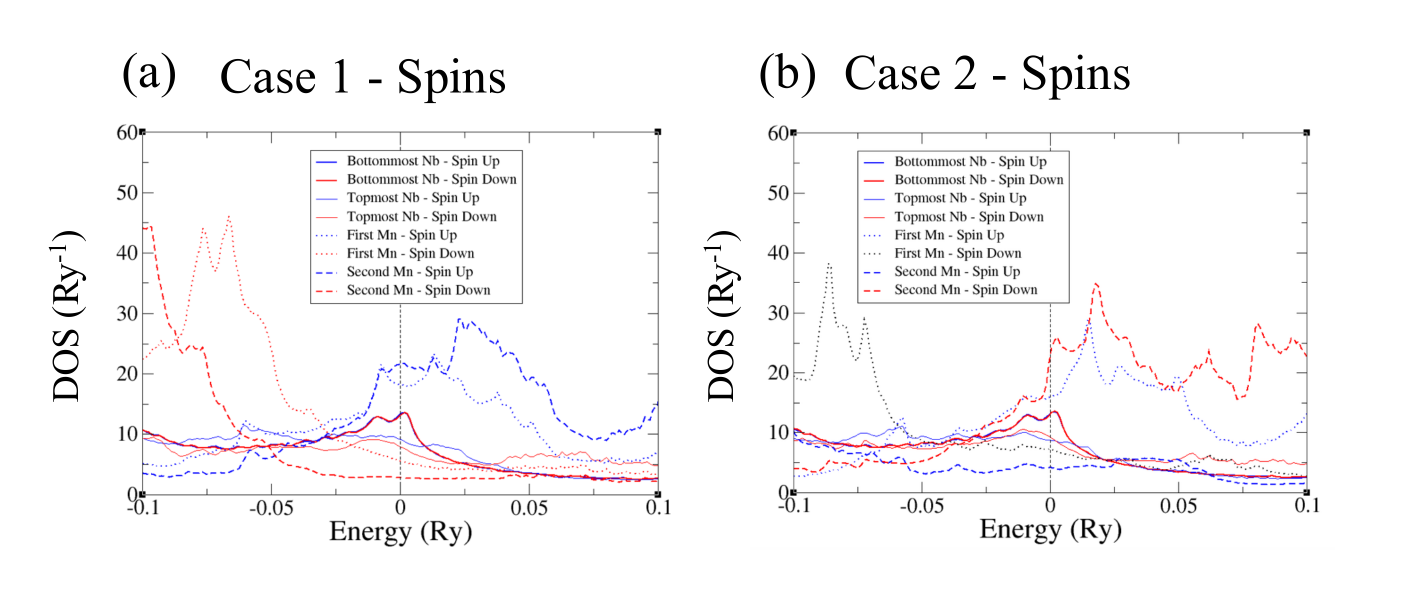}
\caption{Spin decomposition of the normal state LDOS for the double-Mn-layer heterostructure in (a) case 1 - intralayer and interlayer FM ordering, and (b) case 2 - intralayer FM and interlayer AFM ordering, for the bottommost and topmost Nb layers, and the Mn layers.
}
\label{fig:18}
\end{figure*}

In addition, Figure \ref{fig:18} shows the spin decomposition for the intralayer FM cases; Cases 1 and 2, for the aforementioned layers. Again, the bottommost Nb layer shows equal contributions from both spin species, while the topmost Nb shows some exchange splitting which is greater for case 1 (intralayer and interlayer FM ordering) than case 2 (intralayer FM and interlayer AFM ordering) as intuitively expected. 

\section{Bloch Spectral Function (BSF)}

The BSF can be resolved into layer as well as electron-like and hole-like components, which may give more insight into the proximity effect in the systems at hand. A purely electron-like (and symmetrically purely hole-like) BSF would indicate a lack of pairing. However, our calculations trivially show strong pairing on the Nb layers, and not so trivially small, but non-negligible pairing on the Mn layers. The electron-hole symmetry can be reassuringly observed. 

\begin{figure*}
\centering
\includegraphics[width= 0.9 \textwidth]{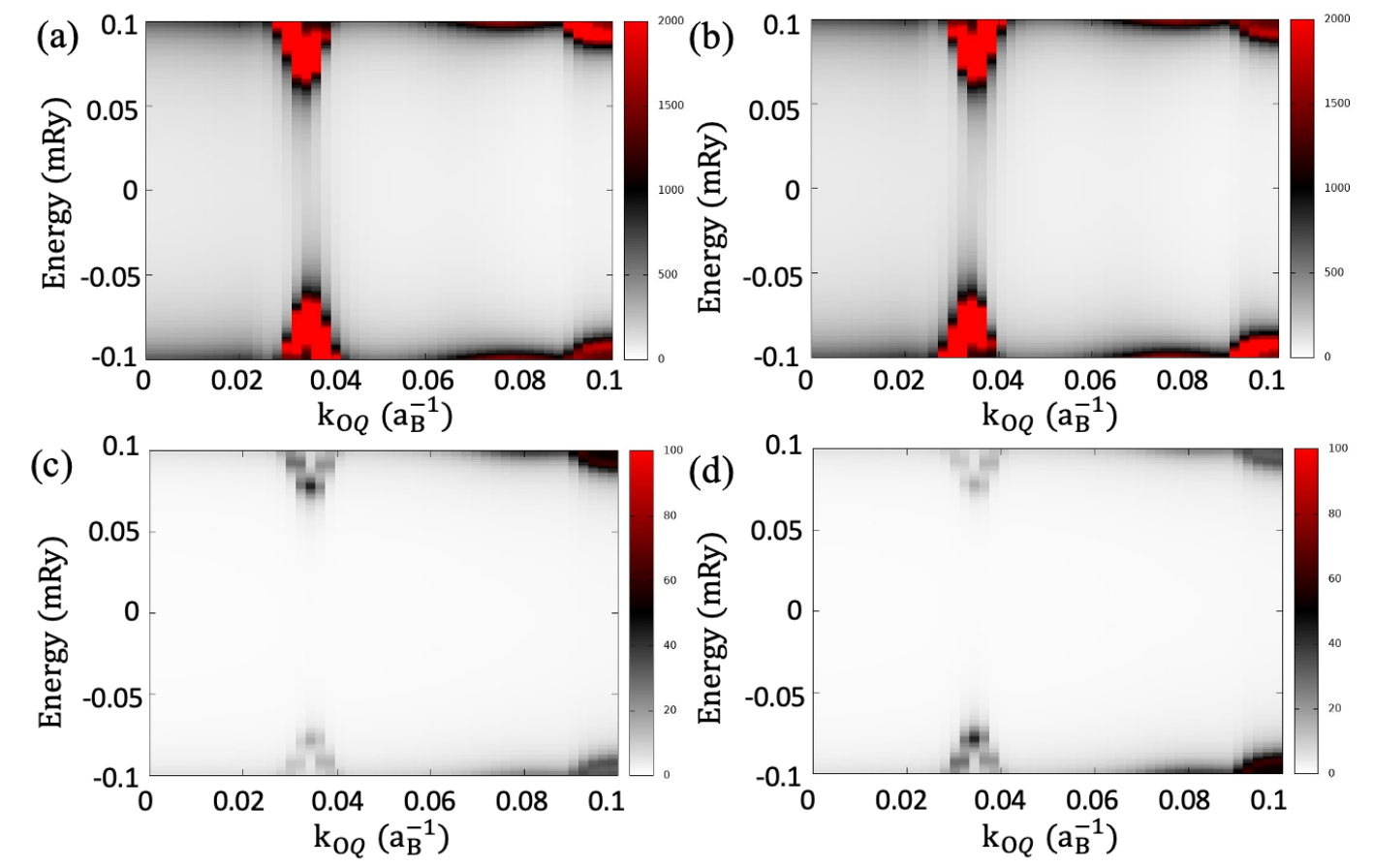}
\caption{BSF for the AFM single-Mn-layer heterostructure along the $k_{\text{OQ}}$ direction between 0 and 0.1 $a_B^{-1}$ showing a clear case for a gap. (a) shows the total electron-like contribution, (b) shows the total hole-like contribution, (c) shows the electron-like contribution in the Mn layer, and (d) shows the hole-like contribution in the Mn layer.}
\label{fig:12}
\end{figure*}

\begin{figure*}
\centering
\includegraphics[width= 0.9 \textwidth]{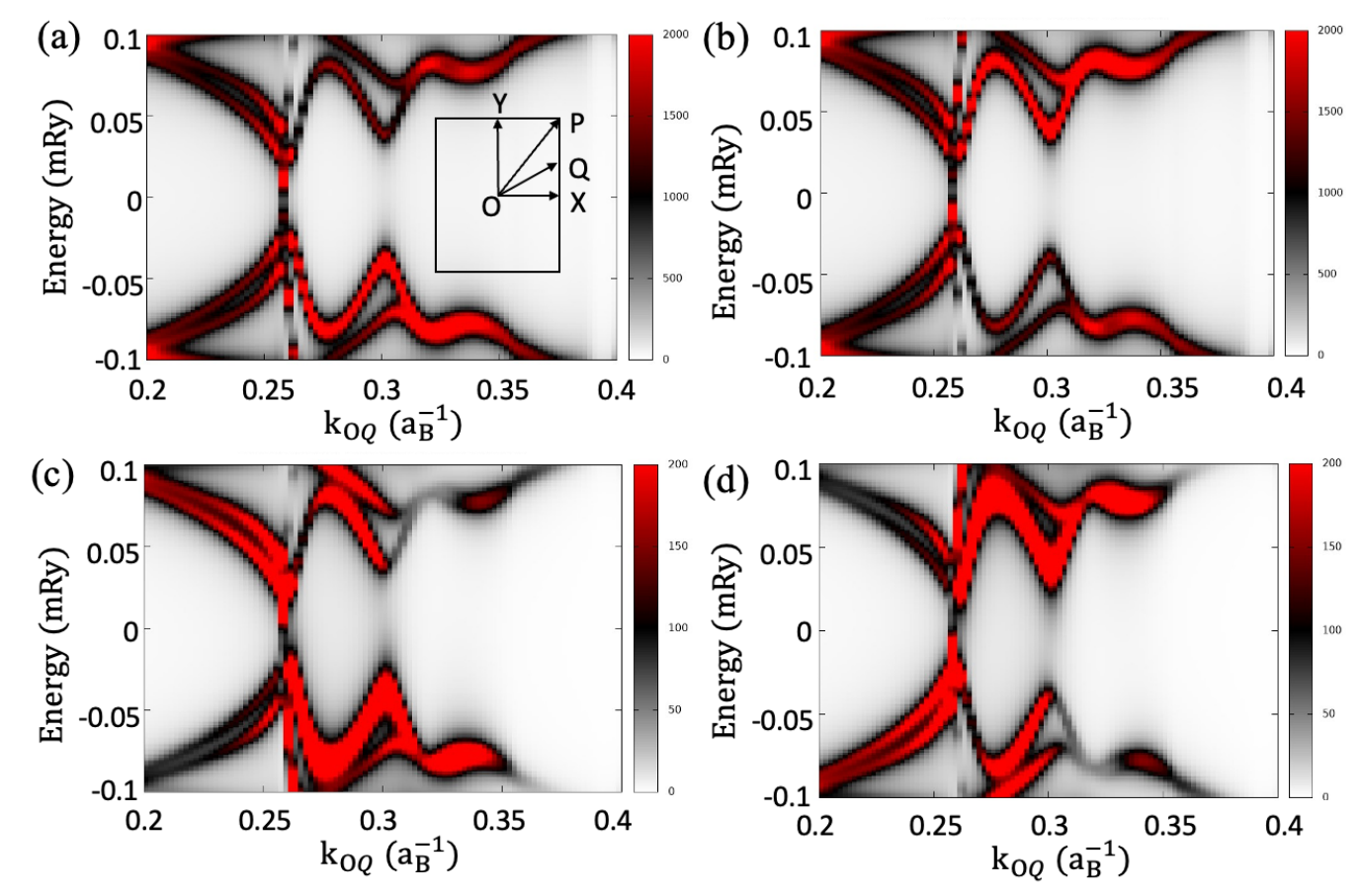}
\caption{BSF for the AFM single-Mn-layer heterostructure along the $k_{\text{OQ}}$ direction between 0.2 and 0.4 $a_B^{-1}$ showing a band crossing the Fermi level at $k_{\text{OQ}} \approx 0.26 \ a_B^{-1}$. (a) shows the total electron-like contribution, (b) shows the total hole-like contribution, (c) shows the electron-like contribution in the Mn layer, and (d) shows the hole-like contribution in the Mn layer.}
\label{fig:13}
\end{figure*}

We start with looking in depth at an example of a clear gap in the single-Mn-layer heterostructure AFM case along the $k_{\text{OQ}}$ direction in $k-$space. Figure \ref{fig:12} shows the total electron-like (a) and hole-like (b) contributions, as well as the electron-like (c) and hole-like (d) contributions specifically for the Mn layer. Next, we show the case that we had to investigate in order to confirm a crossing in the AFM case,  along the $k_{\text{OQ}}$ direction. Figure \ref{fig:13}(a) and (b) again shows the total electron and hole-like contributions, respectively, while Figure \ref{fig:13}(c) and (d) shows the electron and hole-like contributions for the Mn layer. We can clearly see the band crossing the Fermi level at $k_{\text{OQ}} \approx 0.26 \ a_B^{-1}$. 

\begin{figure*}
\centering
\includegraphics[width= 0.9 \textwidth]{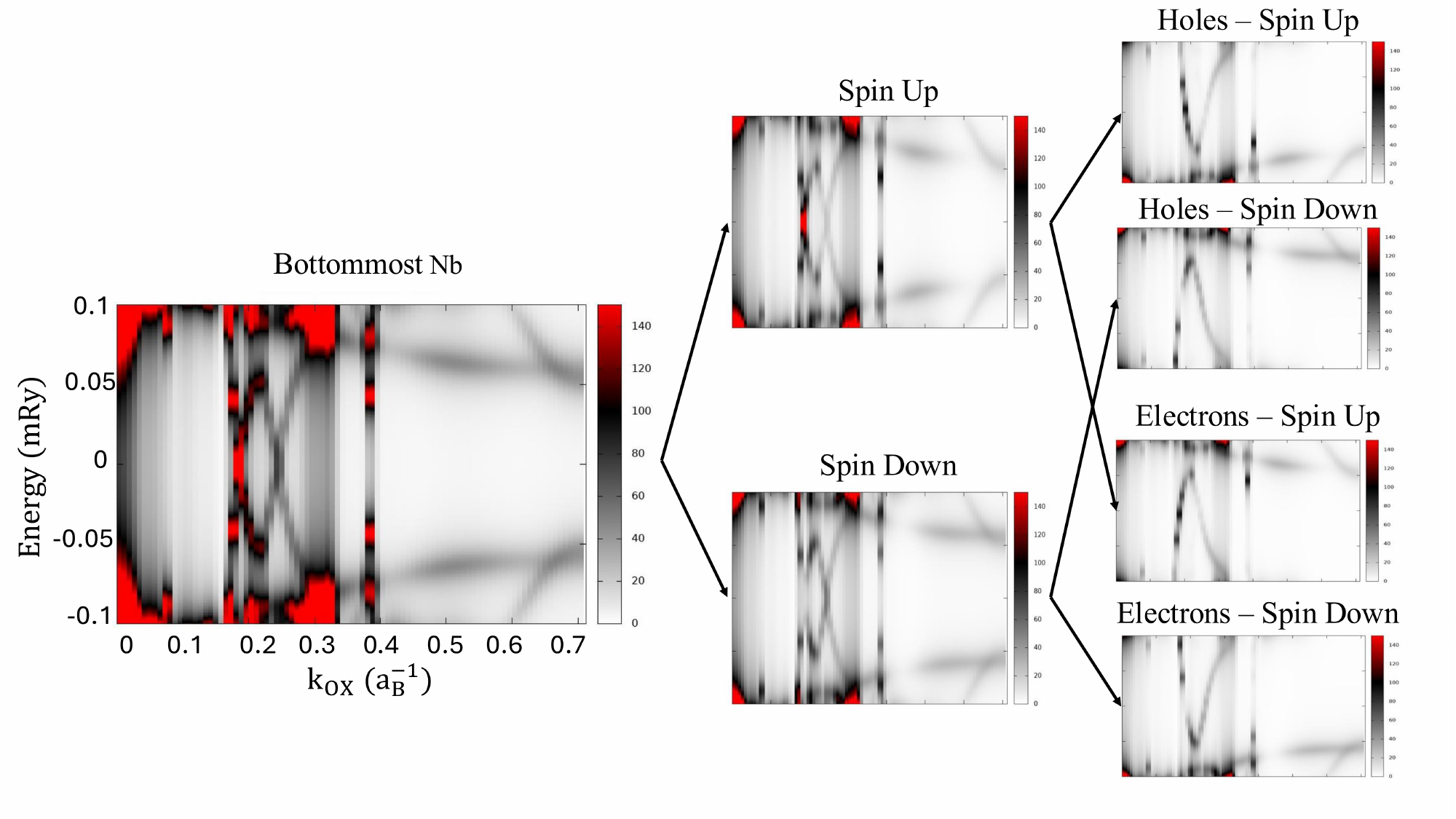}
\caption{BSF for the bottommost Nb layer in the FM single-Mn-layer heterostructure along the $k_{\text{OX}}$ direction showing spin contributions and electron/hole-like contributions, as well as an almost crossing at $k_{\text{OQ}} \approx 0.39 \ a_B^{-1}$}
\label{fig:14}
\end{figure*}

\begin{figure*}
\centering
\includegraphics[width= 0.9 \textwidth]{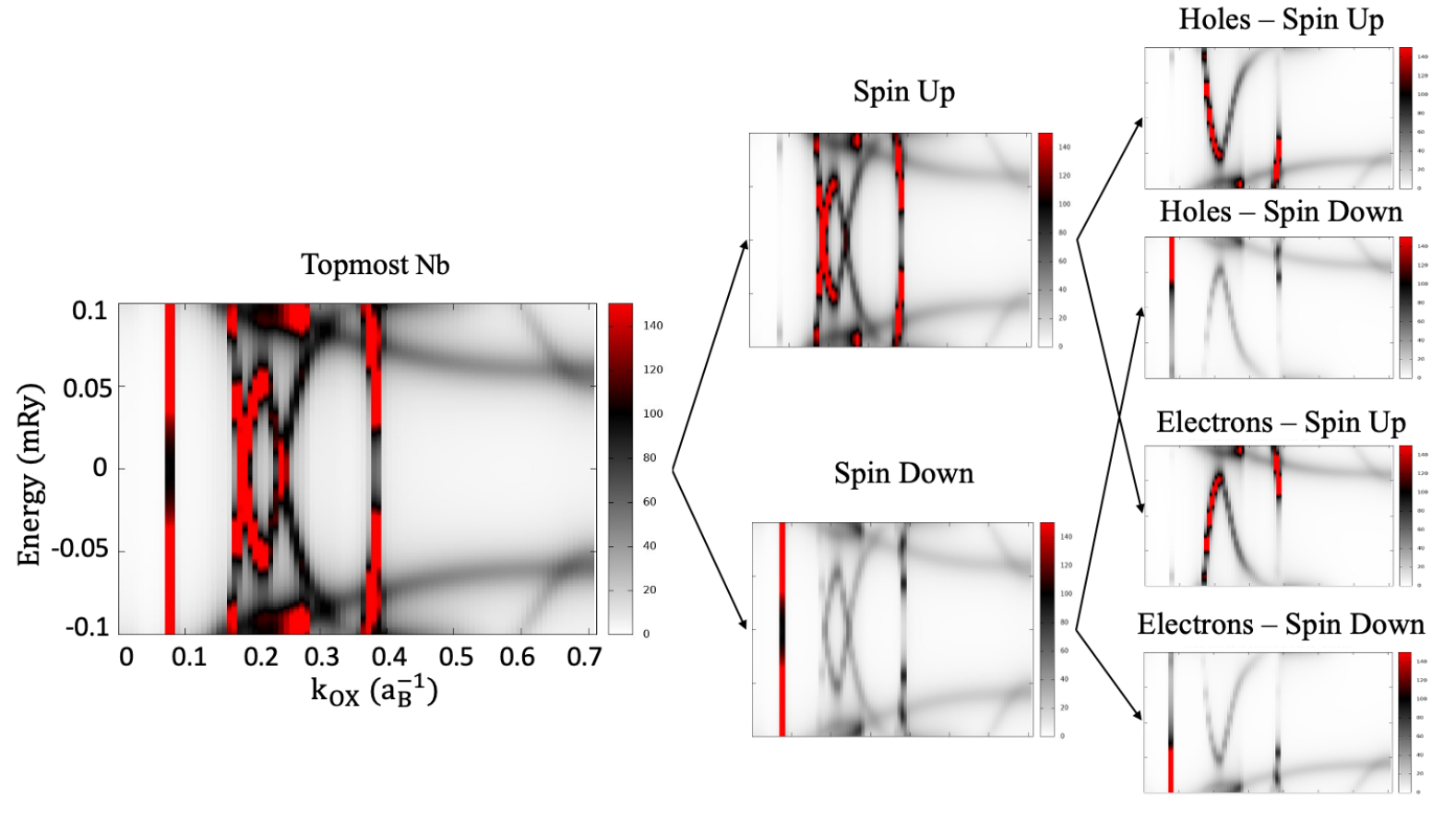}
\caption{BSF for the topmost Nb layer in the FM single-Mn-layer heterostructure along the $k_{\text{OX}}$ direction showing spin contributions and electron/hole-like contributions, as well as an almost crossing at $k_{\text{OQ}} \approx 0.39 \ a_B^{-1}$}
\label{fig:15}
\end{figure*}

\begin{figure*}
\centering
\includegraphics[width= 0.9 \textwidth]{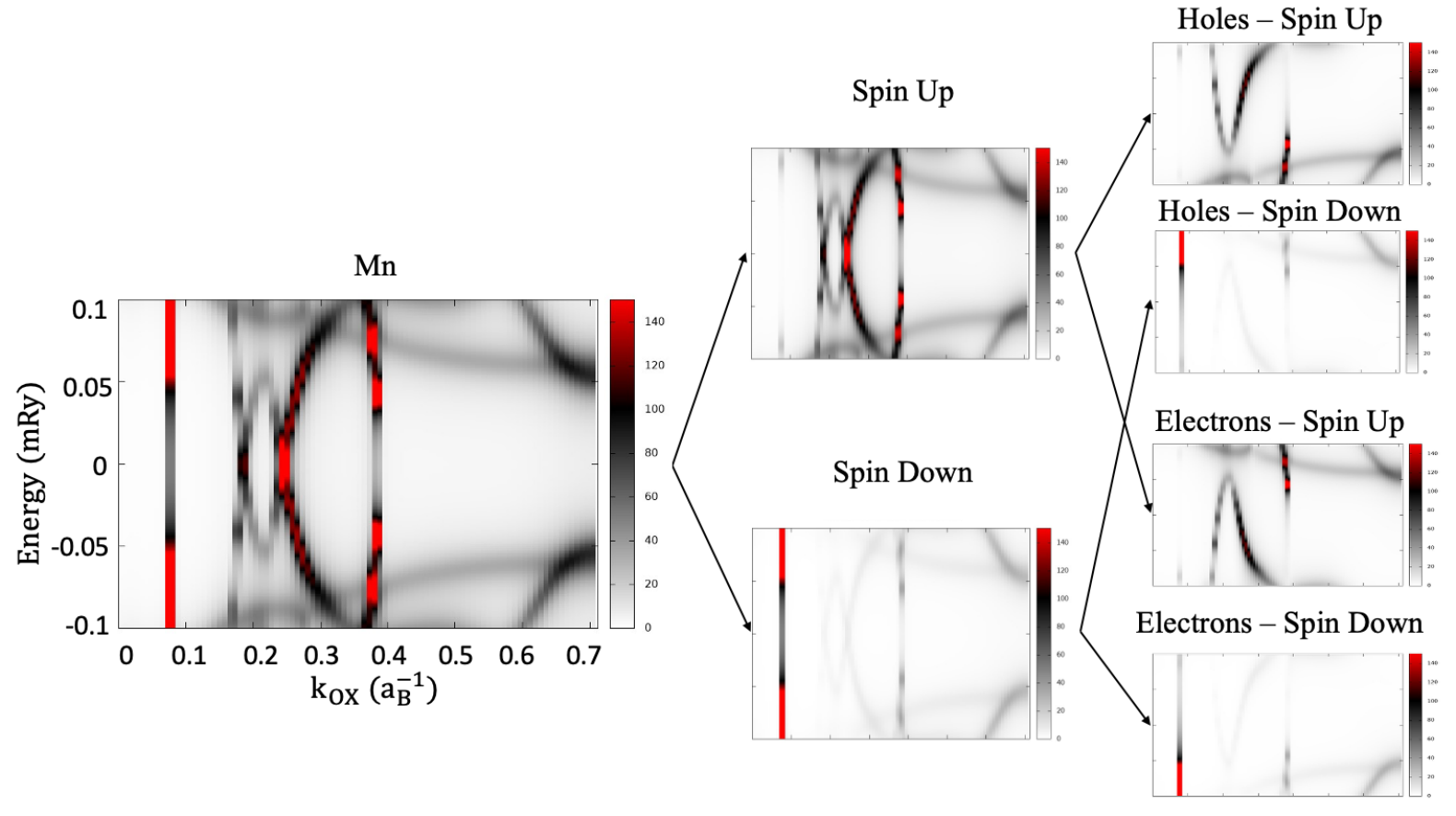}
\caption{BSF for the Mn layer in the FM single-Mn-layer heterostructure along the $k_{\text{OX}}$ direction showing spin contributions and electron/hole-like contributions, as well as an almost crossing at $k_{\text{OQ}} \approx 0.39 \ a_B^{-1}$}
\label{fig:16}
\end{figure*}

For the sake of completeness, we show a case for the single-Mn-layer FM heterostructure along the $k_{\text{OX}}$ direction for three layers; the bottommost Nb in Fig.\ref{fig:14}, the topmost Nb in Fig.\ref{fig:15}, and the Mn layer in Fig.\ref{fig:16}. Here, we show the spin contributions to the BSF and then further decompose into electron and hole-like contributions. We can clearly see the difference between the bottommost and topmost Nb BSF in terms of the spin decomposition since we showed that the induced magnetic moments are only relevant in the two Nb layers closest to the Mn layer and are further negligible. We see in the topmost Nb BSF (Fig.\ref{fig:15}) and the Mn layer BSF (Fig.\ref{fig:16}) that the spin contributions to the Fermi level crossing states are unequal, with the spin down component heavily contributing to the crossing around $k_{\text{OX}} \approx 0.08 \ a_B^{-1}$ and the spin up component heavily contributing to the rest of the crossings at $0.15 \ a_B^{-1} < k_{\text{OX}} < 0.3 \ a_B^{-1}$. In addition, we can also see an example of an almost crossing at $k_{\text{OX}} \approx 0.39 \ a_B^{-1}$ which following the decompositions into spin and electron/hole-like components, we can see is not a real crossing.

\section{Superconducting order Parameter}

In this section, we show in Table \ref{table4} additional information on the components of the superconducting order parameter in the Mn layer/s, for both single and double layer heterostructures; both singlet and triplet (IAT) components, namely, we show the absolute values of each $\chi^{Mn}_{0,\pm{1}}$ and $\chi_S^{Mn}$ in Rydbergs, as well as $\overline{\chi_S^{Mn}}$ which is the ratio between $\chi_S^{Mn}$ and $\chi_S^{Nb}$, bearing in mind that the calculated bulk value of the Nb singlet component $\chi_S^{Nb}$ is $9.7\times10^{-3} \, \text{Ry}$. We can see that the maximum value for $\overline{\chi_S^{Mn}}$ is approximately 4.44\% while the absolute order parameter component values in the Mn layer/s are two or three orders of magnitude less than $\chi_S^{Nb}$.

\begin{table*}
 \caption{Singlet ($\chi_S^{Mn}$) and triplet (IAT) component values ($\chi^{Mn}_{0,\pm{1}}$) in the Mn layer/s (Rydberg), and the ratio of the singlet component value with the Nb bulk value ($\overline{\chi_S^{Mn}}$)\\}
 \begin{tabular}{|c|c|c|c|c|c|}
 \hline
 Magnetic Case & $\chi_S^{Mn}$ & $\overline{\chi_S^{Mn}}$ & $\chi_0^{Mn}$ & $\chi_1^{Mn}$ & $\chi_{-1}^{Mn}$ \\ [1ex]
 \hline\hline
 FM & $4.4\times 10^{-5}$ & 0.0045 & $2.2\times 10^{-4}$ & $1.6 \times 10^{-5}$ & $1.45\times 10^{-4}$ \\
 \hline
 AFM &  $4.3\times 10^{-4}$ & 0.0444 & $2.6\times 10^{-4}$ & $3.55\times 10^{-5}$ & $3.55\times 10^{-5}$ \\
\hline
 Case 1 & \begin{tabular}{@{}c@{}} $3.4\times 10^{-4}$ \\ $-4.7\times 10^{-5}$\end{tabular} & \begin{tabular}{@{}c@{}} 0.0346 \\ -0.0048 \end{tabular} & \begin{tabular}{@{}c@{}} $1.0\times 10^{-4}$ \\ $4.6\times 10^{-5}$ \end{tabular} & \begin{tabular}{@{}c@{}} $4.3\times 10^{-5}$ \\ $1.5\times 10^{-5}$ \end{tabular} & \begin{tabular}{@{}c@{}} $9.5\times 10^{-5}$ \\ $9.3\times 10^{-5}$ \end{tabular} \\
 \hline
 Case 2 & \begin{tabular}{@{}c@{}} $3.1\times 10^{-4}$ \\ $1.7\times 10^{-4}$\end{tabular} & \begin{tabular}{@{}c@{}} 0.0323 \\ 0.0176 \end{tabular} & \begin{tabular}{@{}c@{}} $2.7\times 10^{-4}$ \\ $8.3\times 10^{-5}$ \end{tabular} & \begin{tabular}{@{}c@{}} $3.1\times 10^{-5}$ \\ $6.2\times 10^{-5}$ \end{tabular} & \begin{tabular}{@{}c@{}} $1.1\times 10^{-4}$ \\ $1.0\times 10^{-5}$ \end{tabular} \\ [1ex]
 \hline
 Case 3 & \begin{tabular}{@{}c@{}} $1.84\times 10^{-4}$ \\ $1.77\times 10^{-4}$\end{tabular} & \begin{tabular}{@{}c@{}} 0.01895 \\ 0.0183 \end{tabular} & \begin{tabular}{@{}c@{}} $1.192\times 10^{-4}$ \\ $1.236\times 10^{-4}$ \end{tabular} & \begin{tabular}{@{}c@{}} $6.1\times 10^{-5}$ \\ $3.9\times 10^{-5}$ \end{tabular} & \begin{tabular}{@{}c@{}} $6.1\times 10^{-5}$ \\ $3.9\times 10^{-5}$ \end{tabular} \\ [1ex]
 \hline
 Case 4 & \begin{tabular}{@{}c@{}} $2.6\times 10^{-4}$ \\ $1.3\times 10^{-4}$\end{tabular} & \begin{tabular}{@{}c@{}} 0.0270 \\ 0.0135 \end{tabular} & \begin{tabular}{@{}c@{}} $1.8\times 10^{-4}$ \\ $6.7\times 10^{-5}$ \end{tabular} & \begin{tabular}{@{}c@{}} $2.86\times 10^{-5}$ \\ $7.1\times 10^{-5}$ \end{tabular} & \begin{tabular}{@{}c@{}} $2.86\times 10^{-5}$ \\ $7.1\times 10^{-5}$ \end{tabular} \\ [1ex]
 \hline
\end{tabular}
\label{table4}
\end{table*}